\pdfoutput=1  

\documentclass[aps,prd,superscriptaddress,floatfix,nofootinbib,preprintnumbers,eqsecnum,twocolumn,dvipsnames]{revtex4-1} 

\usepackage{amsmath,float,graphicx,placeins,amssymb,multirow,pgfplots,pgfplotstable}
\usepackage{empheq}
\usepackage{tikz}
\usepackage{tikz-feynman}
\usepackage{subfigure}
\usepackage{comment}
\usepackage{enumerate,slashed,cancel,comment,color,bbm,graphicx,rotating,placeins,mathtools,multirow,hhline}
\usepackage[normalem]{ulem}
\usepackage{array}

\usepackage{hyperref}
\usepackage{color}
\usepackage{xcolor}
 \hypersetup
 {
   colorlinks,
   linkcolor={blue!80!black},
   citecolor={blue!70!black},
   urlcolor={blue!70!black}
 }

\usepackage{lipsum}
\usepackage{diagbox}
\usetikzlibrary{shapes.geometric,arrows}
\graphicspath{}
\usepackage{siunitx}
\usepackage{empheq}

\def\beq{\begin{equation}}
\def\eeq{\end{equation}}
\def\beqn{\begin{eqnarray}}
\def\eeqn{\end{eqnarray}}
\def\half{\mbox{\small ${\frac{1}{2}}$}}

\def\quarter{\mbox{\small ${\frac{1}{4}}$}}

\newcommand{\newc}{\newcommand}
\def\calZ{{\cal Z}}
\def\calM{{\cal M}}
\def\calV{{\cal V}}
\def\calF{{\cal F}}

\def\calH{{\cal H}}

\def\bQ{{\bf Q}}
\def\bT{{\bf T}}

\def\Qs{{\bf q}}

\def\KE{{\rm KE}}
\def\CM{{\rm CM}}

\def\barOmega{{\overline{\Omega}}}
\def\barkappa{{\overline{\kappa}}}

\def\barXi{{\overline{\Xi}}}

\def\half{{\textstyle{1\over 2}}}
\def\quarter{{\textstyle{1\over 4}}}
\def\ie{{\it i.e.}\/}
\def\eg{{\it e.g.}\/}
\def\etc{{\it etc}.\/}
\def\inbar{\,\vrule height1.5ex width.4pt depth0pt}
\def\IR{\relax{\rm I\kern-.18em R}}
 \font\cmss=cmss10 \font\cmsss=cmss10 at 7pt
\def\IQ{\relax{\rm I\kern-.18em Q}}
\def\IZ{\relax\ifmmode\mathchoice
 {\hbox{\cmss Z\kern-.4em Z}}{\hbox{\cmss Z\kern-.4em Z}}
 {\lower.9pt\hbox{\cmsss Z\kern-.4em Z}}
 {\lower1.2pt\hbox{\cmsss Z\kern-.4em Z}}\else{\cmss Z\kern-.4em Z}\fi}

\def\be{\begin{equation}}
\def\ee{\end{equation}}
\def\ba{\begin{eqnarray}}
\def\ea{\end{eqnarray}}

\def\Ncal{\mathcal{N}}

\def\Hcal{\mathcal{H}}

\def\neff {n_{\text{eff}}}

\nonstopmode
\begin{document}

\title{
Betwixt Annihilation and Decay:\\  The Hidden Structure of Cosmological Stasis}

\def\andname{\hspace*{-0.5em}} 

\author{Jonah Barber}
\email[Email address: ]{jbarber2@arizona.edu}
\affiliation{Department of Physics, University of Arizona, Tucson, AZ 85721 USA}

\author{Keith R. Dienes}
\email[Email address: ]{dienes@arizona.edu}
\affiliation{Department of Physics, University of Arizona, Tucson, AZ 85721 USA}
\affiliation{Department of Physics, University of Maryland, College Park, MD 20742 USA}

\author{Brooks Thomas}
\email[Email address: ]{thomasbd@lafayette.edu}
\affiliation{Department of Physics, Lafayette College, Easton, PA  18042 USA}

\begin{abstract}
Stasis is a unique cosmological phenomenon in which the abundances of different energy components in 
the universe (such as matter, radiation, and vacuum energy) each remain fixed even though they scale
differently under cosmological expansion.  Moreover, extended epochs exhibiting stasis are generally 
cosmological attractors in many BSM settings and thus arise naturally and without fine-tuning.  
To date, stasis has been 
found within a number of very different BSM cosmologies.  In some cases, stasis emerges from theories 
that contain large towers of decaying states (such as theories in extra dimensions or string theory).  
By contrast, in other cases, no towers of states are needed, and stasis instead emerges due to thermal 
effects involving particle annihilation rather than decay.  In this paper, we study the dynamics of the 
energy flows  in all of these theories during stasis, and find that these theories all share a common 
energy-flow  structure which in some sense lies between particle decay and particle annihilation.  
This structure has been hidden until now but ultimately lies at the root of the stasis phenomenon, 
with all of the previous stases appearing as different manifestations of this common underlying structure.
This insight not only allows us to understand the emergence of stasis in each of these different scenarios, 
but also provides an important guide for the potential future discovery of stasis in additional cosmological 
systems.
\end{abstract}
\maketitle

\tableofcontents

\def\ie{{\it i.e.}\/}
\def\eg{{\it e.g.}\/}
\def\etc{{\it etc}.\/}
\def\taubar{{\overline{\tau}}}
\def\qbar{{\overline{q}}}
\def\kbar{{\overline{k}}}
\def\bQ{{\bf Q}}
\def\calT{{\cal T}}
\def\calN{{\cal N}}
\def\calF{{\cal F}}
\def\calM{{\cal M}}
\def\calZ{{\cal Z}}

\def\beq{\begin{equation}}
\def\eeq{\end{equation}}
\def\beqn{\begin{eqnarray}}
\def\eeqn{\end{eqnarray}}
\def\apo{\mbox{\small ${\frac{\alpha'}{2}}$}}
\def\half{\mbox{\small ${\frac{1}{2}}$}}
\def\sqapo{\mbox{\tiny $\sqrt{\frac{\alpha'}{2}}$}}
\def\sqap{\mbox{\tiny $\sqrt{{\alpha'}}$}}
\def\sqapxtwo{\mbox{\tiny $\sqrt{2{\alpha'}}$}}
\def\aptwo{\mbox{\tiny ${\frac{\alpha'}{2}}$}}
\def\apofour{\mbox{\tiny ${\frac{\alpha'}{4}}$}}
\def\bosqtwo{\mbox{\tiny ${\frac{\beta}{\sqrt{2}}}$}}
\def\btosqtwo{\mbox{\tiny ${\frac{\tilde{\beta}}{\sqrt{2}}}$}}
\def\apofour{\mbox{\tiny ${\frac{\alpha'}{4}}$}}
\def\sqaptwo{\mbox{\tiny $\sqrt{\frac{\alpha'}{2}}$}  }
\def\apoeight{\mbox{\tiny ${\frac{\alpha'}{8}}$}}
\def\sapoeight{\mbox{\tiny ${\frac{\sqrt{\alpha'}}{8}}$}}

\newc{\gsim}{\lower.7ex\hbox{{\mbox{$\;\stackrel{\textstyle>}{\sim}\;$}}}}
\newc{\lsim}{\lower.7ex\hbox{{\mbox{$\;\stackrel{\textstyle<}{\sim}\;$}}}}
\def\calM{{\cal M}}
\def\calV{{\cal V}}
\def\calF{{\cal F}}
\def\bQ{{\bf Q}}
\def\bT{{\bf T}}
\def\Qs{{\bf q}}

\def\half{{\textstyle{1\over 2}}}
\def\quarter{{\textstyle{1\over 4}}}
\def\ie{{\it i.e.}\/}
\def\eg{{\it e.g.}\/}
\def\etc{{\it etc}.\/}
\def\inbar{\,\vrule height1.5ex width.4pt depth0pt}
\def\IR{\relax{\rm I\kern-.18em R}}
 \font\cmss=cmss10 \font\cmsss=cmss10 at 7pt
\def\IQ{\relax{\rm I\kern-.18em Q}}
\def\IZ{\relax\ifmmode\mathchoice
 {\hbox{\cmss Z\kern-.4em Z}}{\hbox{\cmss Z\kern-.4em Z}}
 {\lower.9pt\hbox{\cmsss Z\kern-.4em Z}}
 {\lower1.2pt\hbox{\cmsss Z\kern-.4em Z}}\else{\cmss Z\kern-.4em Z}\fi}



\section{Introduction}


It has long been known that we live in an expanding universe.  This, in turn, has a major impact on the 
composition of the universe.  Since the different forms of energy in the universe (such as vacuum energy, 
matter, and radiation) scale differently under cosmological expansion, even the relative amounts of these 
different energy components present in the universe are constantly in flux.  It is therefore somewhat 
surprising that many models of physics beyond the Standard Model (BSM) actually lead to something different, 
specifically a phenomenon wherein the cosmological abundances of the different energy components actually 
remain constant despite cosmological expansion.  This unexpected phenomenon has been dubbed
``stasis''~\cite{Dienes:2021woi}, and such stasis epochs can generally extend across many $e$-folds of 
cosmological expansion.

Such stasis epochs turn out to be fairly ubiquitous, arising in many different BSM cosmologies.  Many of 
these cosmologies involve theories that have large towers of states.  These include, for example, theories 
involving extra spacetime dimensions, in which the towers of states correspond to the different 
Kaluza-Klein (KK) modes of the theory~\cite{Dienes:2021woi,Dienes:2023ziv,Dienes:2024wnu}.   However, 
these also include string theories, in which the towers of states correspond to the different string 
excitations, and even include strongly coupled gauge theories, in which the towers of states correspond 
to the different bound-state resonances~\cite{Dienes:2021woi,Dienes:2023ziv}.   Stasis has even been found 
in theories in which the states in such towers are primordial black holes (PBHs) with different 
masses~\cite{Barrow:1991dn, Dienes:2022zgd}.  Indeed, the stasis phenomenon emerges within these sorts 
of tower-based theories regardless of whether their mass spectra grow 
polynomially~\cite{Dienes:2021woi, Barrow:1991dn, Dienes:2022zgd, Dienes:2023ziv, Dienes:2024wnu} or 
even exponentially~\cite{Halverson:2024oir}.

Stasis has also been found in theories that lack such towers.  For example, stasis has been shown to 
arise in thermal theories in which only a single matter species and a single radiation species are present~\cite{Barber:2024vui,Barber_toappear_model}.  This demonstrates that stasis extends even into the 
thermal domain.

Needless to say, the stasis phenomenon gives rise to a host of new theoretical possibilities across the 
entire cosmological timeline.  For example, like other theories of BSM physics, the stasis phenomenon 
leads to the possibility of highly non-traditional cosmological epochs and correspondingly non-standard 
expansion histories~\cite{Batell:2024dsi}.  However, within the BSM theories outlined above,
a stasis epoch is not only ubiquitous but also a dynamical attractor.  As a result, a stasis epoch is 
essentially unavoidable in such BSM cosmologies.  This then adds extra importance to the task of 
understanding the physical consequences and signatures of such an epoch, ranging from potential 
implications for primordial density perturbations, dark-matter production, and structure formation all 
the way to early reheating, early matter-dominated eras, and even the age of the universe.   
There has even been a proposal for a new kind of inflation --- a so-called {\it stasis inflation} --- in 
which the inflationary era is itself a stasis epoch~\cite{Dienes:2024wnu}.

In each of these different kinds of stasis, the cosmological abundances of the different energy components 
are stable because the effects of cosmological expansion are balanced against processes that transfer energy 
from one component to another.  For example, in stases involving matter and radiation, the effects of 
cosmological expansion tend to cause the abundance of radiation to decrease while the abundance of matter 
increases.  This occurs because the energy density associated with radiation dilutes more rapidly than that 
of matter in an expanding universe.  However, if the matter can decay back into radiation, this may provide 
a counter-balancing effect that enables the abundances of matter and radiation to remain constant.  
In general, we shall use the word ``pump'' to refer to any such process that gives rise to such a 
counter-balancing transfer of energy, since its operation tends to mitigate the natural effects of 
cosmological expansion and restore the balance between the different abundances involved.  At first glance, 
such a balancing may appear to be fine-tuned, especially since a purely cosmological effect such as 
cosmological expansion is being balanced against a pump effect coming from purely standard particle-physics.
However, the fact that the stasis phenomenon is an attractor --- indeed, even a {\it global}\/ 
attractor --- guarantees that such a balancing will come into existence even if the universe did not begin 
in such a balanced state.

That said, the different stases discussed in the prior literature emerge within the cosmologies corresponding 
to wholly different BSM models.  Some stases rely on the existence of large (or even infinite) towers of 
states, while others rely on only a single state species.  Likewise, these stases rely on the effects 
produced by a variety of different pumps.  Some stases rely on the decay of matter into radiation, as 
discussed above, while others instead rely on the {\it annihilation}\/ of matter into radiation --- a process 
which has a completely different dependence on the existing matter energy density.  Some 
stases~\cite{Dienes:2023ziv} even rely on the existence of {\it phase transitions}\/ that convert vacuum 
energy into matter (such as occur in axion misalignment production).  Moreover, some of these stases are 
completely non-thermal, with pumps that are completely independent of temperature, while others are 
intrinsically thermal, with pumps that depend on the temperature of certain states in the theory.

Despite the rich variety of stasis models, the ubiquity of stasis suggests that something deeper is at 
play --- that each of these different stases is related to the others in a fundamental way.  Indeed, one 
suspects that each of our different stases may be a viewed as a different manifestation of a single underlying 
phenomenon.  This paper is devoted to addressing this question.   Indeed, as we shall find, these stases 
ultimately all share a common energy-flow structure that corresponds to a pump lying somewhere between 
particle decay and particle annihilation.  This structure has been hidden until now, but we shall demonstrate 
that this structure  ultimately lies at the root of the stasis phenomenon, with all of the previous stases 
appearing as different manifestations of this common underlying structure.


\section{Preliminaries}


In this section, we begin by establishing the general algebraic underpinnings of the stasis phenomenon.
We also discuss the various possibilities for energy flow in models that realize stasis, and then concentrate 
on one particular flow pattern which we shall eventually find to be central to all instances of the stasis 
phenomenon.

\subsection{General considerations \label{sec:general_considerations}}

Our treatment begins in the same manner as in previous analyses of the stasis phenomenon, only generalized 
to the case of arbitrary numbers of energy components with arbitrary equations of state.
Towards this end, we begin, as in Refs.~\cite{Dienes:2021woi, Dienes:2023ziv}, by assuming a flat 
Friedmann-Robertson-Walker (FRW) universe containing a set of different energy components labeled by the 
index $i$, each with an energy density $\rho_i$ and a corresponding  fixed equation-of-state parameter $w_i$.  

For convenience we shall order these energy components in terms of increasing $w_i$, such that 
$w_i<w_{i+1}<w_{i+2}<...$.  We shall also define their corresponding  abundances 
\beq
  \Omega_i   ~\equiv~ \frac{8\pi G}{3H^2} \,\rho_i~,~
  \label{Omegadef}
\eeq
where $H$ is the Hubble parameter and $G$ is Newton's constant.  From Eq.~(\ref{Omegadef}) it follows that
\beq
  \frac{d\Omega_i}{dt} ~=~ \frac{8\pi G}{3} 
    \left( \frac{1}{H^2} \frac{d\rho_i}{dt} - 2 \frac{\rho_i}{H^3} \frac{dH}{dt} \right)~.
\label{stepone}
\eeq
The final term in this expression can be evaluated through the use of the Friedmann ``acceleration'' 
equation for $dH/dt$.  In general, this equation takes the form
\beqn
  \frac{dH}{dt} ~&=&~ -H^2 - \frac{4\pi G}{3} \left(  \sum_i \rho_i + 3 \sum_i p_i\right) \nonumber\\
  ~&=&~ - \frac{3}{2} H^2 \left(1+ \sum_i \Omega_i w_i \right)~.
  \label{acceleq}
\eeqn
Note that we may also parametrize the time-evolution of the Hubble parameter via the general relation
\beq
  \frac{dH}{dt} ~=~ -\frac{3}{\kappa}  H^2 ~,
  \label{kappadef}
\eeq
whereupon comparison with Eq.~(\ref{acceleq}) yields
\beq
  \kappa ~=~ \frac{2}{1+ \sum_i w_i \Omega_i} ~.
\label{kappadef2} 
\eeq
We shall find that many of our results simplify when expressed in terms of $\kappa$.  Indeed, 
substituting Eq.~(\ref{kappadef}) into Eq.~(\ref{stepone})  yields
\beq
  \frac{d\Omega_i}{dt} ~=~ \frac{8\pi G}{3H^2} \, \frac{d\rho_i}{dt}  +  \frac{6 H \Omega_i}{\kappa}~.
  \label{dOmdt}
\eeq
These are thus general relations for the time-evolution of the abundances $\Omega_i$ in an arbitrary 
flat universe.  Note that in general, $\rho_i$, $\Omega_i$, and $\kappa$ are all time-dependent quantities.

Given these relations, our final step is to insert appropriate ``equations of motion'' for the various 
$d\rho_i/dt$ into Eq.~(\ref{dOmdt}).  In general, we shall assume that these equations of motion take 
the general form
\beq
  \frac{d\rho_i}{dt} ~=~ -3(1+w_i) H \rho_i  
    + \sum_{j<i} P_{ji}^{(\rho)} - \sum_{j>i} P_{ij}^{(\rho)} ~.
  \label{eoms}
\eeq
The first term on the right side of this equation represents the general redshifting effect that arises 
due to cosmological expansion.  By contrast, the final two terms account for possible sources and sinks 
amongst the different energy components.  In general, these sources and sinks are expressed in terms of 
various ``pumps'' $P^{(\rho)}_{ij}$ which conserve energy but change the {\it distribution}\/ of this energy 
density amongst the different energy components. We shall adopt the convention that the pump $P^{(\rho)}_{ij}$ 
describes the rate at which energy density  is transferred from component $i$ to component $j$.  Thus, with 
this convention, the second term in Eq.~(\ref{eoms}) represents the {\it contributions}\/ to $\rho_i$ from 
energy components with {\it smaller}\/ equation-of-state parameters, while the third term in Eq.~(\ref{eoms}) 
represents the {\it loss}\/ of energy density $\rho_i$ to energy components with {\it larger}\/ equation-of-state
parameters.  We have chosen these explicit forms for these two terms in Eq.~(\ref{eoms}) based on our expectation 
that our pumps will generally transfer energy density in the direction of increasing equation-of-state parameters.
However, if this ever fails to be the case for a pair of components $i$ and $j$, this simply means that the 
corresponding pump is negative.  Thus, in all cases, the expression in Eq.~(\ref{eoms}) remains completely 
general for all possible pumps and energy transfers regardless of their signs.

Substituting Eq.~(\ref{eoms}) into Eq.~(\ref{dOmdt}), we then find
\beq
  \frac{d\Omega_i}{dt} ~=~  \left[ \frac{6}{\kappa}  - 3 (1+w_i) \right] H\Omega_i
    +\sum_{j<i} P_{ji} - \sum_{j>i} P_{ij}~
  \label{convert3}
\eeq
where
\beq
  P_{ij} ~\equiv~ \frac{8\pi G}{3 H^2} \, P_{ij}^{(\rho)}~.
  \label{abundance_pump}
\eeq
Indeed, while $P_{ij}^{(\rho)}$ serves as a pump for the transfer of {\it energy densities}\/ $\rho(t)$, 
we may think of $P_{ij}$ as a corresponding pump for the transfer of {\it abundances}\/ $\Omega(t)$.

In general, we are seeking a steady-state ``stasis'' solution in which the abundances $\Omega_i$ all take 
constant values $\barOmega_i$.  Clearly such a solution will arise if the effect of the cosmological 
expansion is precisely counterbalanced by the effect of the pumps.  We therefore wish to impose, 
at the very minimum, the conditions that $d\Omega_i/dt=0$.  Given Eq.~(\ref{convert3}), we thus see that 
our general conditions for stasis are given by
\beq 
  \boxed{~~\sum_{j>i} P_{ij} - \sum_{j<i} P_{ji} ~=~ 
    \left[ \frac{6}{\barkappa}  - 3 (1+w_i) \right] H \,\barOmega_i~~}
  \label{pregenstasisconditions}
\eeq
where $\barkappa$ is the stasis value of $\kappa$ and where $H$ now represents the Hubble parameter during 
stasis.  However, during stasis, we can actually solve Eq.~(\ref{kappadef}) to find that 
$H(t) = \barkappa / (3t)$.  Thus, our condition in Eq.~(\ref{pregenstasisconditions}) becomes
\beq 
  \boxed{~~\sum_{j>i} P_{ij} - \sum_{j<i} P_{ji} ~=~ \biggl[ 2-(1+w_i) \barkappa\biggr] 
    \, \barOmega_i\, \frac{1}{t}~.~~}
  \label{condition}
\eeq
We thus have a separate stasis condition for each energy component $i$ in the universe, except that 
one of these constraint equations is the sum of the others.  This latter degeneracy reflects the fact that 
$\sum_i \Omega_i=1$, or equivalently that $\sum_i d\Omega_i/dt=0$.  Of course, our derivation of 
Eq.~(\ref{condition}) has assumed that each of the energy components is individually in stasis.  This 
assumption is implicit in our assumption that $\kappa$ takes a constant value during stasis.
  
Given the result in Eq.~(\ref{condition}), it then follows that the pumps which produce a stasis state 
must all share a common scaling behavior
\beq
  \sum_{j>i} P_{ij} - \sum_{j<i} P_{ji} ~\sim~ \frac{1}{t}~
  \label{genstasisconditions}
\eeq
during stasis.
Indeed, this behavior must hold for each $i$.
 This in turn implies that we must equivalently have      
\beq
  \sum_{j>i} P^{(\rho)}_{ij} - \sum_{j<i} P^{(\rho)}_{ji} ~\sim~ \frac{1}{t^3}~
  \label{genstasisconditions2}
\eeq
during stasis.   Indeed, so long as these  scaling relations hold, both sides of Eq.~(\ref{condition}) 
will behave identically as functions of time.  It then follows that if Eq.~(\ref{condition}) is satisfied 
at one instant, it will continue to be satisfied for all times, resulting in a true stasis solution which 
extends across many $e$-folds of cosmological expansion until some other aspect of our original setup 
changes.  These issues will be discussed extensively below.

As an example of the above results, let us consider a universe with three energy components:  
matter $M$ with $w= 0$, radiation $\gamma$ with $w= 1/3$, and vacuum energy $\Lambda$ with $w$ taken to 
be fixed at a value $w_\Lambda$ (which we can assume is close to  $-1$).  We therefore generally have 
three independent pumps --- $P_{\Lambda M}$, $P_{\Lambda \gamma}$, and $P_{M\gamma}$ --- and from 
Eq.~(\ref{condition}) we obtain the corresponding stasis conditions~\cite{Dienes:2023ziv}
\beqn
  P_{\Lambda M} + P_{\Lambda \gamma}  ~&=&~  \Bigl[2 - (1+w_\Lambda)\barkappa\Bigr]~ 
    \barOmega_\Lambda \, \frac{1}{t} ~~\nonumber\\
  P_{M \gamma} - P_{\Lambda M}  ~&=&~ \Bigl[ 2-\barkappa \Bigr] ~\barOmega_M\, \frac{1}{t}
    \nonumber\\
  -P_{\Lambda \gamma} - P_{M \gamma} ~&=&~ \!\left[ 2-\frac{4\barkappa}{3} \right]~ 
    \barOmega_\gamma \, \frac{1}{t}~.~
  \label{eq:alstruc}
\eeqn

\subsection{Generalized pumps}

Thus far, our discussion has largely followed that in Refs.~\cite{Dienes:2021woi,Dienes:2023ziv}, only 
generalized for an arbitrary number of energy components.  However, the resulting dynamics of our system 
depends critically on the specific pumps $P^{(\rho)}$ and $P$ that appear in Eqs.~(\ref{eoms}) 
and~(\ref{convert3}).  We shall therefore now turn our attention to the pumps and the manner in which they 
affect the resulting stasis dynamics.  Several different pumps were considered in previous
works~\cite{Dienes:2021woi,Dienes:2022zgd,Dienes:2023ziv,Barber:2024vui}.  Indeed, in each case the pumps 
that were utilized were motivated by the particular particle-physics model of stasis under study.   
However, these disparate choices tended to obscure the commonalities that underlay their resulting stases.  

One of the goals of this paper is to expose this common underlying structure.  Towards this end, we shall 
now contemplate a wider class of pumps than have previously been considered.  However, it turns out that 
the primary features associated with this underlying structure are largely independent of the number or 
specific identities of the distinct energy components.  Therefore, for simplicity, we shall henceforth 
restrict our attention to a minimal system consisting of only two energy components.  

A two-component system has only one pump, and this pump is determined by the particular system under study. 
For example, if our system involves a transfer of energy density from matter to radiation which arises 
through the decay of a matter field $\phi$, the equations of motion for the matter and radiation energy 
densities in Eq.~(\ref{eoms}) will incorporate a pump of the form $P_{M\gamma}^{(\rho)}\sim \Gamma \rho_M$ 
where $\Gamma$ is the corresponding decay width.  Likewise, if the energy-density transfer is realized through 
an instantaneous phase transition (such as a scalar field $\phi$ transitioning from an overdamped to 
underdamped phase, as in Ref.~\cite{Dienes:2023ziv}), we will again have a pump which can be modeled as linear 
in the energy density $\rho_M$.   Indeed, these linear pumps are precisely the sorts of pumps which
have been examined in Refs.~\cite{Dienes:2021woi, Barrow:1991dn, Dienes:2022zgd,Dienes:2023ziv}.

However, we can also contemplate other, more exotic means of energy-density transfer.  For example, if this 
transfer occurs through the two-body {\it annihilation}\/ (rather than decay) of a matter field $\phi$ 
into radiation, the corresponding pump would take the {\it quadratic}\/ form $P_{M\gamma}^{(\rho)}\sim \rho_M^2$
where the overall coefficient might involve an annihilation cross-section.  Likewise, higher-order scattering 
processes could yield pumps involving higher powers of $\rho_M$.

With these examples in mind, in this paper we shall consider a general pump of the form
\beq
  P_{ij}^{(\rho)} ~=~ Z \,\rho_i^n
  \label{genn}
\eeq
where $(i,j)$ refer to our two energy components (which might well be matter and radiation), 
where $Z$ is a positive prefactor which is independent of $\rho_i$, and where $n$ is an arbitrary 
exponent.  We then find that our corresponding abundance pump $P_{ij}$ in Eq.~(\ref{abundance_pump}) 
is given by
\beq
  P_{ij} ~=~ \frac{8\pi G}{3H^2} \, Z \, \rho_i^n ~=~
    Z \, \left( \frac{8\pi G}{3H^2}\right)^{1-n} \Omega_i^n~.
  \label{pumpy}
\eeq
Our stasis condition in Eq.~(\ref{condition}) then takes the form
\beq
  Z \, \barOmega_i^{n-1} ~=~ \left( \frac{8\pi G}{3}\right)^{n-1}  
    \left[ \frac{6}{\barkappa} - 3(1+w_i)\right]\,  H^{3-2n} ~.~
\label{stasiscondition2}
\eeq
For example, in the case of a matter/radiation stasis, we have the condition
\beq
  Z \, \barOmega_M^{n-1} ~=~ \left(\frac{8\pi G}{3}\right)^{n-1} 
    (1-\barOmega_M)\,  H^{3-2n} ~.
  \label{stasiscondition3}
\eeq

Given the forms of these stasis conditions, we immediately identify two values of $n$ which may be of interest.  
The first, of course, is $n\,{=}\,1$.  Indeed, for $n\,{=}\,1$ we lose all factors of $8\pi G/3$. However, 
we immediately see from Eq.~(\ref{stasiscondition3}) that we can never achieve a stasis solution with a fixed, 
time-independent abundance $\barOmega_M$ for $n\,{=}\,1$ unless $Z$ itself has a $1/t$ scaling behavior under 
cosmological expansion.  It is highly non-trivial to arrange models for which this occurs.

Of course, the models of Refs.~\cite{Dienes:2021woi,Barrow:1991dn,Dienes:2022zgd,Dienes:2023ziv} do exhibit 
matter/radiation stasis with $n\,{=}\,1$.  However, they accomplish this in an entirely different manner, 
namely by partitioning $\Omega_M$ into a large set of sub-abundances corresponding to individual matter 
subcomponents $\phi_\ell$ and then introducing a time-dependence into this partition.  Fortunately, such 
abundance distributions with time-dependent partitions are not unnatural or fine-tuned, and arise in many 
well-known theories of physics beyond the Standard Model. Thus the models of 
Refs.~\cite{Dienes:2021woi, Barrow:1991dn,Dienes:2022zgd,Dienes:2023ziv} are compelling in their own rights, 
and in Sect.~\ref{sec:n32tower} we shall show how those models actually ``secretly'' fit within the larger 
framework that is the subject of this paper. 

That said, our interest at this juncture is in straightforward stasis scenarios with only a single component 
$\phi$, and we have seen that single-component pumps with $n\,{=}\,1$ cannot achieve stasis in this manner.

\subsection{Stasis with $n=3/2$ \label{sec:threehalftheory} }

There is, however, another interesting value of $n$ within Eqs.~(\ref{stasiscondition2}) 
and~(\ref{stasiscondition3}): $n\,{=}\,3/2$.  Indeed, for $n\,{=}\,3/2$, the Hubble parameter drops out 
of our stasis conditions!  Focusing on the case of matter/radiation stasis which will form the centerpiece 
of this paper, and further assuming that $Z$ is a constant during stasis, we then find that for 
$n\,{=}\, 3/2$ our stasis condition in Eq.~(\ref{stasiscondition3}) reduces to the simple form
\beq
  ~\sqrt{\,\barOmega_M}  ~=~   Y \left(1-\barOmega_M\right)~
  \label{threehalfstasiscondition}
\eeq
where the constant $Y$ is given by
\beq 
  Y ~\equiv~ \sqrt{ \frac{8\pi G}{3 Z^2}}~.
  \label{Ydef}
\eeq

At first glance, it may seem that a non-integer value of $n$ such as $n=3/2$ cannot emerge from any underlying 
self-consistent model of particle physics.   However, as we shall demonstrate in future sections of this paper, 
this value of $n=3/2$ actually emerges in scenarios which give rise to stasis  in a rather interesting and 
compelling way, along with a particle-physics realization for $Z$ according to which $Z$ remains fixed under 
cosmological expansion.  Indeed, we shall find that these particle-physics realizations are as well motivated 
as those in Refs.~\cite{Dienes:2021woi,Barrow:1991dn,Dienes:2022zgd,Dienes:2023ziv}.  We shall therefore 
temporarily accept the legitimacy of the $n=3/2$ choice and proceed to explore the properties of the stasis 
that emerges.

With $n=3/2$, our stasis condition in Eq.~(\ref{threehalfstasiscondition}) has only one unknown variable 
$\barOmega_M$.  Thus, as long as $Z$ is invariant under cosmological expansion, we can actually solve for 
our stasis abundances directly from this equation alone, obtaining 
\beq 
  \barOmega_M= \frac{ \sqrt{1+4Y^2}-1}
    {\sqrt{1+4Y^2}+1}~,~~~~ \barOmega_\gamma = \frac{2}{\sqrt{1+4Y^2} +1}~.~
  \label{stasisabundances}
\eeq
Indeed, our ability to deduce the  stasis abundances in this simple manner is yet another critical difference 
relative to previous stasis analyses involving multiple matter subcomponents. 

These solutions for the stasis abundances make intuitive sense.  As we tune our pump to become increasingly 
strong by taking $Z$ increasingly large (or $Y$ increasingly small), we find that $\barOmega_\gamma$ becomes 
increasingly large while $\barOmega_M$ becomes increasingly small.  This demonstrates the increased 
effectiveness of the corresponding conversion of matter into radiation, despite the effects of cosmological 
expansion.  Indeed, as $Z\to \infty$, we see that $(\barOmega_M,\barOmega_\gamma)\to (0,1)$ (which may be 
regarded as a case of stasis with essentially full conversion), while in the opposite limit with $Z\to 0$ 
we find $(\barOmega_M,\barOmega_\gamma)\to (1,0)$.  These results are consistent with our expectation that 
$0\leq \barOmega_{M,\gamma} \leq 1$  as $Z$ is varied between $0$ and $\infty$.  Interestingly, however, 
we observe that we obtain a stasis solution for our abundances {\it for any value of $Z$}\/.  Thus, no 
matter how strong or weak our pump may be, there exists a stasis solution in which the effects of cosmological 
expansion are precisely counterbalanced against the effects of the pump!  In this sense we see that stasis 
is a generic property of this system.

\begin{figure*}[t!]
   \centering
   \includegraphics[keepaspectratio, width=0.52\textwidth]{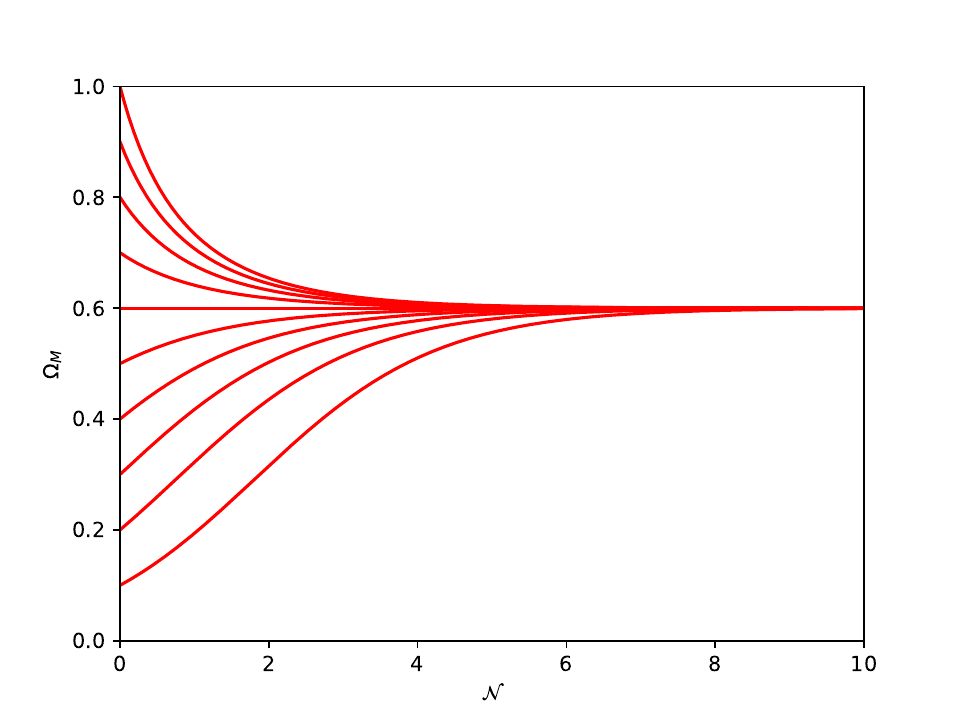}
\hskip -0.35 truein
  \includegraphics[keepaspectratio, width=0.52\textwidth]{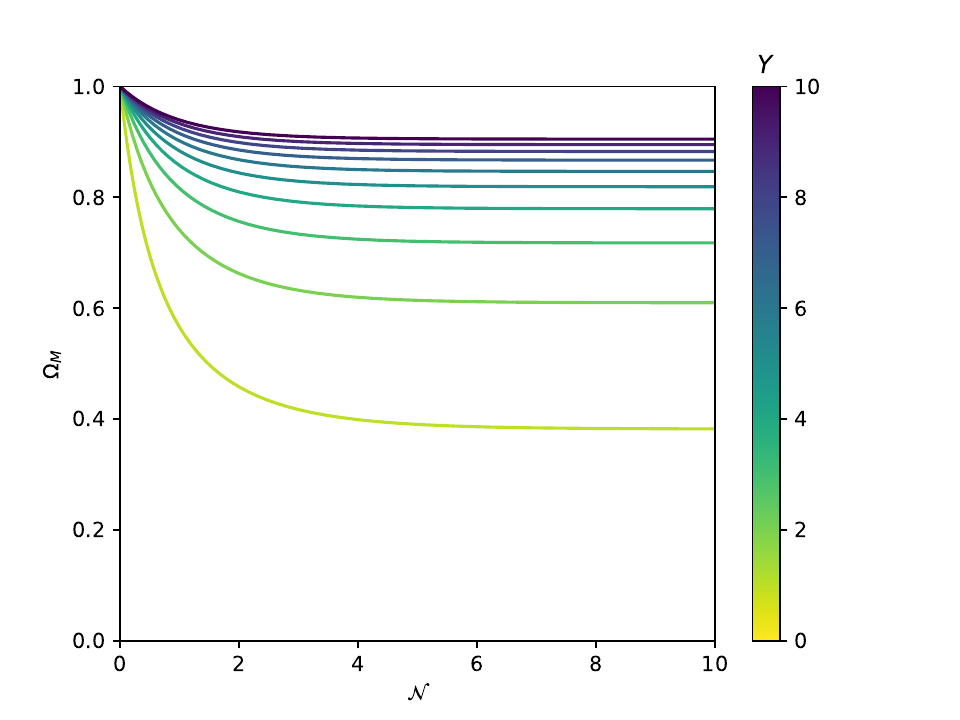}
  \caption{Stasis with $n=3/2$.  
  {\it Left panel}\/: The matter abundances $\Omega_M(t)$ for different initial values 
  $\Omega_M^{(0)}$ at the initial time $t=t^{(0)}$, plotted as functions of the number $\calN$ of 
  elapsed $e$-folds since $t^{(0)}$.  For each curve we have taken $n=3/2$ and chosen a benchmark value 
  of $Z$ such that $\barOmega_M=0.6$.  The different curves shown, bottom to top, correspond to initial 
  values $\Omega_M^{(0)}=n/10$ for increasing integers $n$ within the range $1\leq n\leq 10$.
  {\it Right panel}\/:   Here we plot the matter abundances $\Omega_M(t)$ for different values of $Y$, 
  as defined in Eq.~(\ref{Ydef}), holding $n=3/2$ and $\Omega_M^{(0)}=1$ fixed for all curves.  The 
  different curves shown, bottom to top, correspond to increasing integer values of $Y$ within the 
  range $1\leq Y\leq 10$.   The results in both panels illustrate that our system for $n=3/2$ evolves 
  towards a fixed stasis value $\barOmega_M$ regardless of changes in initial conditions, in accordance 
  with the global-attractor nature of the $n=3/2$ stasis solution.
\label{fig:3-2stasis}}
 \end{figure*}

When $\Omega_M\not= \barOmega_M$, our system is not in stasis.  Indeed, $\Omega_M$ then evolves according 
to Eq.~(\ref{convert3}), which for $n=3/2$ takes the simple form
\beq
  \frac{d\Omega_M}{dt} ~=~ \frac{ H\Omega_M}{Y}
    \left[ -\sqrt{\Omega_M} + Y (1-\Omega_M)\right]~.
  \label{eq:dWdt_with_Y}
\eeq
However, given the stasis abundances in Eq.~(\ref{stasisabundances}), we can rewrite this equation in the form
\beq
  \frac{d\Omega_M}{dt} ~=~ \frac{H \Omega_M}{Y} \left[ 
    \sqrt{\,\barOmega_M}-\sqrt{\Omega_M}       + Y \left( \barOmega_M -\Omega_M\right)\right]~.
  \label{attractor}
\eeq
Since $H$ and $Y$ are necessarily positive, we see that $d\Omega_M/dt >0 $ if $\Omega_M<\barOmega_M$, while
$d\Omega_M/dt <0$ if $\Omega_M> \barOmega_M$.  Thus, if $\Omega_M\not= \barOmega_M$, we find that $\Omega_M$ 
always evolves towards the stasis solution for which $\Omega_M=\barOmega_M$.   Indeed, this is true for all
$0\leq \Omega_M\leq 1$.   Thus our stasis solution is in fact a {\it global attractor}\/ for this system.  
Even if our system does not start in stasis, it will inevitably enter into stasis at a later time unless 
there is a later change to the model itself.

We also note that if we define the number $\calN$ of cosmological $e$-folds which have elapsed since the 
original production time $t^{(0)}$ via 
\beq
  \calN ~\equiv~ \log\left[ \frac{a(t)}{a(t^{(0)})} \right]~,
\eeq
we then have $d\calN/dt=H$.   
We may then rewrite Eq.~(\ref{eq:dWdt_with_Y}) in the form
\beq
  \frac{d\Omega_M}{d\calN} ~=~ -\frac{\Omega_M^{3/2}}{Y} + \Omega_M  (1-\Omega_M)~.
  \label{dWdN_thf_one}
\eeq
Notably, this removes all dependence on an explicit time variable from the right side of this equation, 
thereby demonstrating that this dynamical system is autonomous.  This in turn guarantees that the behavior 
of our system is mathematically independent of the particular choice of initial time at which the dynamical 
evolution is chosen to begin.

In Fig.~\ref{fig:3-2stasis} we illustrate the $n=3/2$ stasis and its global-attractor behavior.  In each 
panel, we plot the matter abundances $\Omega_M(t)$ as functions of the number of elapsed $e$-folds since 
the initial production time $t^{(0)}$, taking $n=3/2$ in all cases but varying the initial conditions of 
this time-evolution.  In each case, we see that our abundances evolve towards the stasis values 
$\barOmega_M$ given in Eq.~(\ref{stasisabundances}), thereby numerically verifying the global-attractor 
behavior of the $n=3/2$ stasis solution.

At first glance, it might seem counter-intuitive that this $n=3/2$ stasis can generally persist across 
large numbers of $e$-folds, given that we have only one decaying matter field.  However, as we shall see, 
this stasis will in fact be  infinite in duration unless other changes in the background cosmology intercede. 

Of more pressing concern, however, is the choice $n=3/2$ that underlies this stasis.  Such a fractional 
value does not have an immediate particle-physics interpretation, and it may not seem possible to realize 
such a fractional power of $\rho$ within a pump term in traditional models of  particle physics.  One 
might also worry that $n$ must be {\it fine-tuned}\/ to have the value $3/2$ in order to obtain the 
behavior illustrated in Fig.~\ref{fig:3-2stasis}.~ Indeed, one can verify that the behavior illustrated 
in Fig.~\ref{fig:3-2stasis} holds only for $n=3/2$ and is significantly disturbed if $n$ deviates even 
slightly --- either positively or negatively --- away from this value.   However, as we shall demonstrate 
in Sects.~\ref{sec:tower} and \ref{sec:thermal}, these issues ultimately have a common resolution.

In fact, we shall take this one step further.   Thus far, we have found that the choice $n=3/2$ leads to 
a stasis.  However, we now claim that {\it this is actually the fundamental stasis that underlies all of 
the realizations of stasis previously considered in the literature.}\/  In other words, we assert that 
each of the stases previously examined in the literature is ``secretly'' an $n=3/2$ stasis.

We are thus left with a number of questions.  How can we find evidence of $n=3/2$ behavior within each of 
the stases which have previously been studied?  Indeed, how is it that each of these systems --- which 
apparently have $n=1$ or $n=2$ --- nevertheless manages to exhibit this $n=3/2$ behavior?  Moreover, we would 
like to understand how this $n=3/2$ scaling behavior actually emerges {\it dynamically}\/ from our systems 
involving particle decays, underdamping transitions, or thermal effects.  Along the way, we would also like 
to understand the role played by the tower in the cases which are apparently $n=1$, or the role of the thermal 
effects in the cases which are apparently $n=2$.    
 
The rest of this paper is devoted to answering these questions.  In Sect.~\ref{sec:tower}, we shall address 
these questions for the apparent $n=1$ stases involving towers.  Indeed,  we shall find that the existence of 
the tower secretly ``deforms'' the value $n=1$ into an effective value $3/2$.  The precise manner in which 
this deformation occurs will be discussed in detail below.  Likewise, in Sect.~\ref{sec:thermal}, we shall 
consider the apparent $n=2$ thermal stasis and demonstrate that the thermal effects induce a similar 
deformation in the other direction, shifting $n=2$ to the same effective value $3/2$.  Finally, in
Sect.~\ref{sec:extracting}, we shall push these steps one step further and show how we can actually 
extract a hidden underlying $n=3/2$ theory from each of these cases.

\subsection{Relation to pump time-scaling\label{relationtopumptimescaling}}

Before plunging into these analyses, however, we make one final observation.  In general, $n$ is 
defined implicitly through Eq.~(\ref{genn}).   However, motivated by this relation, in this paper we 
will define an effective value of $n$ more generally as
\beq
  n_{\rm eff}~\equiv~ \frac{ d \log P_{ij}^{(\rho)}}{d \log \rho_i}~
  \label{ndef1}
\eeq
where $\rho_i$ and $P_{ij}^{(\rho)}$ are the total matter energy density and the energy-density pump 
respectively, where our pump transfers energy density from energy component $i$ to energy component $j$.
Note that this definition for $n_{\rm eff}$ can be equivalently written in the forms
\beq
  n_{\rm eff} ~=~ \frac{\partial_t \log P_{ij}^{(\rho)}}{\partial_t \log \rho_i} 
    ~=~ \frac{\partial_\Ncal \log P_{ij}^{(\rho)}}{\partial_\Ncal \log \rho_i}~.
  \label{ndef2}
\eeq
Indeed, while the definitions in Eqs.~(\ref{ndef1}) and (\ref{ndef2}) reproduce the values of $n$ that would 
emerge from Eq.~(\ref{genn}) in simple cases (such as those in which $Z$ is a constant), they also allow 
$n_{\rm eff}$ to take non-integer values and most importantly to {\it evolve}\/ dynamically as a function 
of time as our system approaches stasis.  It is therefore these latter definitions of $n_{\rm eff}$ that we 
shall adopt in the following.

However, these definitions have an important implication.  During stasis, the time-independence of 
$\Omega_i$ implies that
\beq
  \partial_t \log\rho_i ~=~ -\frac{2}{t}~.
  \label{denom0}
\eeq
Likewise, if we assume that our abundance pump $P_{ij}(t)$ scales during stasis as $t^p$ 
where $p$ is an arbitrary scaling exponent, we find that
\beq
  \partial_t \log P_{ij}^{(\rho)} ~=~ \frac{p-2}{t}~.
  \label{numer0}
\eeq
Thus, dividing Eq.~(\ref{numer0}) by Eq.~(\ref{denom0}), we find that during stasis $n_{\rm eff}$ must 
be related to $p$ via
\beq
  n_{\rm eff} ~=~ 1-\frac{p}{2}~.
  \label{neffprelation}
\eeq
With $n_{\rm eff}=3/2$ we thus have $p=-1$, which is consistent with
Eq.~(\ref{genstasisconditions}).

This result indicates that $n_{\rm eff}$ and the exponent $p$ are directly related to each other during 
stasis.  However, these variables convey very different things.  In general, $p$ describes how our pumps 
scale with time regardless of how they are built in terms of the fundamental fields in our theory.
By contrast, $n_{\rm eff}$ carries information about whether the energy-exchange process which constitutes 
the pump might be a decay process, a two-body annihilation process, or something else.  Moreover, these two 
variables are not even directly related to each other in a universal way --- indeed, it is only during stasis 
that they are related via Eq.~(\ref{neffprelation}).   At all other times, $n_{\rm eff}$ depends not only 
on the time-scaling of our pump $P$ but also on the time-scaling of the Hubble parameter $H$ as well as on 
the time-scaling of the abundance $\Omega_i$, both of which are in principle unknown.  Thus, in general, 
$n_{\rm eff}$ carries more information about the overall behavior of our system than does $p$. 

In this paper, we shall therefore concentrate on $n_{\rm eff}$ and its evolution.   Indeed, as discussed 
above, our interest is in  understanding how our cosmological systems --- systems which manifestly have 
$n=1$ or $n=2$ pumps --- nevertheless evolve to have $n_{\rm eff}=3/2$ during stasis.  This of course 
{\it incorporates}\/ the evolution of $p$ towards the value $-1$, but also incorporates a number of other 
evolving scaling relations as well.  This is ultimately why Eq.~(\ref{neffprelation}) holds only in the 
stasis limit, but not along any of the {\it approaches}\/ to stasis that we shall be studying.


\section{A new look at tower-based stases:   How $n=1$ pumps produce $n_{\rm eff}=3/2$ scaling behavior\label{sec:tower}}


In this section we shall discuss how the $n=1$ pumps that are involved in the tower-based stases of 
Refs.~\cite{Dienes:2021woi,Dienes:2023ziv} manage to produce  $n_{\rm eff}=3/2$ scaling behaviors during stasis.
We shall begin by focusing on the matter/radiation stasis of Refs.~\cite{Dienes:2021woi,Dienes:2023ziv}.
This stasis is apparently an $n=1$ stasis, utilizing a pump of the form in Eq.~(\ref{genn}) with $n=1$.  
However, as we shall review, stasis is achieved in this scenario by partitioning $\Omega_M$ amongst an entire 
tower of component states $\phi_\ell$, with $\ell=0,1,...,N-1$ where $N\gg 1$.  We shall then describe a new 
approach to understanding this stasis, one which rests upon the formulation of a new tower-based 
``level-shift'' symmetry that underlies the stasis phenomenon in this scenario.  As we shall see, this 
symmetry is essentially a reflection of the self-similarity of our system as time proceeds and as the 
transitions work their way down the tower.  Utilizing this symmetry, we will then demonstrate how the 
value of $n_{\rm eff}$ is deformed by the dynamics of the system, and how the specific value $n_{\rm eff}=3/2$ 
ultimately emerges within this stasis.  Finally, we shall then discuss the other tower-based stases of
Ref.~\cite{Dienes:2023ziv}, focusing on the vacuum-energy/matter stasis as an example.  As we shall see, 
this stasis behaves similarly to the matter/radiation stasis and ultimately also yields an $n_{\rm eff}=3/2$ 
scaling behavior. Indeed, this value of $n_{\rm eff}$ occurs through the effects of a vacuum-energy/matter 
version of the same tower-based level-shift symmetry.  In general, we consider this level-shift self-similarity 
symmetry to be at the root of the stasis phenomenon in all tower-based stases.

\subsection{Review of tower-based matter/radiation stasis \label{sect:towerbasedreview}}

Motivated by many models of BSM physics, the fundamental premise of the tower-based matter/radiation 
stasis in Refs.~\cite{Dienes:2021woi,Dienes:2023ziv} is that the matter sector of our theory comprises 
a large tower of individual matter components $\phi_\ell$.  These components are presumed to have 
corresponding masses $m_\ell$, energy densities $\rho_\ell$, and abundances $\Omega_\ell$.  We thus have 
$\rho_M=\sum_\ell \rho_\ell$ and $\Omega_M=\sum_\ell \Omega_\ell$.  Each of the $\phi_\ell$  components 
is unstable and decays to radiation with decay width $\Gamma_\ell$.  The equations of motion for these 
components then take the form
\beq
  \frac{d\rho_\ell}{dt} ~=~ 
    -3H\rho_\ell - \Gamma_\ell \rho_\ell~,
  \label{prepump31}
\eeq
whereupon we see that
\beq
  \frac{d\rho_M}{dt} ~=~
    \sum_\ell  \frac{d\rho_\ell}{dt} ~=~ 
    -3H\rho_M  - \sum_\ell \Gamma_\ell \rho_\ell~.
\eeq
The corresponding equation for radiation takes the form
\beq
  \frac{d\rho_\gamma}{dt} ~=~
    -4H\rho_M  + \sum_\ell \Gamma_\ell \rho_\ell~.
  \label{prepump34}
\eeq
Comparing with Eq.~(\ref{eoms}), we thus see that our total energy-density pump from matter to radiation 
takes the form
\beq
  P^{(\rho)}_{M\gamma} ~=~ \sum_\ell \Gamma_\ell \rho_\ell~,
  \label{pumpavgGamma}
\eeq 
which leads to the corresponding total abundance pump
\beq
  P_{M\gamma} ~=~ \sum_\ell \Gamma_\ell \,\Omega_\ell ~.
  \label{pumpavgGamma2}
\eeq 
In other words, $P_{M\gamma}$ is proportional to the abundance-weighted average of the decay widths across 
the tower:
\beq
  P_{M\gamma} ~=~ \langle \Gamma \rangle \,\Omega_M ~
\label{pumpavgGamma3}
\eeq
where
\beq
  \langle \Gamma \rangle ~\equiv ~ \displaystyle{
    \frac{ \sum_\ell \Gamma_\ell \Omega_\ell }   {\sum_\ell \Omega_\ell}} ~.
\eeq

We thus see that our pump ultimately depends on the behaviors of the abundances $\Omega_\ell$ and decay 
widths $\Gamma_\ell$ across the tower.  Once again taking motivation from various models of BSM physics, 
we assume that these follow the general scaling relations~\cite{Dienes:2021woi} 
\beqn 
  \Omega_\ell^{(0)} &=& \Omega_0^{(0)} \left(\frac{m_\ell}{m_0}\right)^\alpha~\nonumber\\
    \Gamma_\ell &=& \Gamma_0 \left(\frac{m_\ell}{m_0}\right)^\gamma~
  \label{scalings1}
\eeqn
where
\beq 
  m_\ell ~=~   m_0 + (\Delta m) \,\ell^\delta~.
  \label{scalings2}
\eeq
In these relations, $\lbrace \alpha,\gamma,\delta\rbrace$ are general scaling exponents;  
$\lbrace m_0, \Delta m, \delta\rbrace$ are all positive; and the `$(0)$' superscripts indicate that 
the corresponding quantities are to be evaluated at the time $t^{(0)}$ at which the abundances are 
initially established.  In general, as discussed in Refs.~\cite{Dienes:2021woi,Dienes:2023ziv}, the 
scaling exponents $\lbrace \alpha,\gamma,\delta\rbrace$ depend on the particular BSM model under study.
Similarly, $\lbrace \Omega_0^{(0)},\Gamma_0, m_0, \Delta m\rbrace$ are further model-dependent parameters.

Cosmological expansion in this mixed matter/radiation universe tends to induce each matter abundance 
$\Omega_\ell$ to grow.  This growth persists until approximately $\tau_\ell\equiv 1/\Gamma_\ell$, at 
which point the decay of $\phi_\ell$ becomes significant and $\Omega_\ell$ begins to fall exponentially.
Indeed, during stasis, each individual $\Omega_\ell(t)$ is given by~\cite{Dienes:2021woi,Dienes:2023ziv}
\beqn
  \Omega_\ell(t) ~=~ \Omega_\ell^{\ast}  
    \left( \frac{t}{t_\ast}\right)^{2-\barkappa}  e^{-\Gamma_\ell (t-t^{(0)})}~
  \label{MGOmegal}
\eeqn
where $t_\ast$ is some early fiducial time within this stasis epoch, where $\overline{\kappa}$ is the stasis 
value of the general quantity $\kappa$ in Eq.~(\ref{kappadef2}), and where 
$\Omega_\ell^{\ast}\equiv \Omega_\ell(t_\ast)$.  Within Eq.~(\ref{MGOmegal}), the factor scaling as a power 
of $(t/t_\ast)$ represents the power-law growth of the abundances due to cosmological expansion while the final 
exponential factor is the result of $\phi_\ell$ decay.  Of course, since $\gamma>0$, the more massive a state 
is, the more rapidly it decays.  Thus the states at the top of the tower decay first, then the next-highest 
states, and so forth down the tower.  Meanwhile, while a given state within the tower is decaying, the 
lighter states within the tower have abundances that are still continuing to grow. This process then continues 
until the bottom state within the tower is reached and decays.

What is remarkable is that the {\it sum}\/ of these time-dependent matter abundances $\Omega_\ell(t)$ within 
the tower ultimately evolves towards a certain value $\barOmega_M$ and then remains essentially {\it fixed}\/ 
at that value throughout this decaying process.  In other words, this system quickly begins to exhibit a 
{\it stasis}\/  between the total matter and radiation abundances, with each remaining essentially fixed over 
an extended period stretching across many $e$-folds of cosmological expansion.  Furthermore, such a stasis 
arises for {\it any}\/ values of the parameters appearing in Eqs.~(\ref{scalings1}) and (\ref{scalings2}) 
provided that~\cite{Dienes:2021woi}
\beq
  0 ~<~ \eta ~\leq ~ \frac{\gamma}{2}~
\eeq
where 
\beq
  \eta ~\equiv~ \alpha + \frac{1}{\delta}~.
  \label{etadef}
\eeq 
Indeed, in all such cases the corresponding
stasis value for $\kappa$ in Eqs.~(\ref{kappadef}) and (\ref{kappadef2}) is given by~\cite{Dienes:2021woi}
\beq
  \barkappa~=~ 2- \frac{\eta }{\gamma}   ~,
  \label{stasiskappa}
\eeq
whereupon we obtain the stasis matter abundance~\cite{Dienes:2021woi}
\beq 
  \barOmega_M ~=~ 4- \frac{6}{\barkappa} ~=~
  \frac{2\gamma - 4 \eta}{2\gamma - \eta} ~.
\eeq 
This behavior is illustrated in Fig.~2 of Ref.~\cite{Dienes:2021woi}, where the parameters chosen 
yield a stasis lasting approximately 15 $e$-folds.  Note that once stasis is reached, the individual 
$\Omega_\ell(t)$ curves are densely overlapping, with an effective envelope function scaling as 
$t^{1/(\gamma\delta)}$.

In this section, we have discussed the case of matter/radiation stasis where the transition between matter 
and radiation is effected through particle decay.   However, as we shall see, all tower-based stases are of 
this general type, with pump-induced transitions proceeding down (or occasionally 
{\it up}~\cite{Dienes:2022zgd}) a tower of states.  In this case, the corresponding pump is given in
Eqs.~(\ref{pumpavgGamma}) through~(\ref{pumpavgGamma3}).   As we see upon comparison between Eq.~(\ref{genn}) 
and~(\ref{pumpavgGamma}),  this is an example of an $n=1$ pump --- \ie, a pump $P_{M\gamma}^{(\rho)}$ which 
is {\it linear}\/ in the corresponding energy density $\rho_M$.

This example also furnishes us with an understanding of why it is necessary to have a tower of states for 
such a pump.  If there had been only one state decaying, the pumps described in Eqs.~(\ref{pumpavgGamma2}) 
and (\ref{pumpavgGamma3}) would reduce to simply 
\beq
  P_{M\gamma}~=~ \Gamma \,\Omega_M
  \label{singlestate}
\eeq
where $\Gamma$ is the decay width of this single state.  Thus, during stasis, such a pump would necessarily be 
constant.  However, we have already seen in Eq.~(\ref{genstasisconditions}) that any pump during stasis must 
scale with time as $1/t$.  Thus a single state experiencing an $n=1$ pump cannot give rise to stasis.  By contrast, 
partitioning this total abundance $\Omega_M$ into individual constituent contributions $\Omega_\ell$ from a 
tower of states allows for the possibility of achieving an overall $1/t$ scaling for the pump, since the 
{\it distribution}\/ of the total abundance across the constituents may be time-dependent and thereby carry 
the $1/t$ scaling even while the total $\Omega_M$ remains constant.  Indeed, by direct calculation one 
finds (see, \eg, Eq.~(3.14) of Ref.~\cite{Dienes:2021woi}) that
\beq
  \langle \Gamma \rangle ~=~ \left(\frac{\eta}{\gamma} \right) \, \frac{1}{t} ~.
  \label{bracketGamma}
\eeq
This result holds completely generally, even though each individual constituent decay width $\Gamma_\ell$ 
is a constant, and does not assume stasis.   Thus, when our system is not in stasis, we see from 
Eq.~(\ref{pumpavgGamma3}) that the pump $P_{M\gamma}$ fails to scale as $1/t$ only because $\Omega_M$ 
carries its own additional time dependence.   By contrast, when our system enters stasis,  $\Omega_M$ 
becomes constant.   Our pump then scales as $1/t$ because of the relation in Eq.~(\ref{bracketGamma}).

\subsection{Self-similarity and level-shift symmetries}

The above description of a tower-based stasis essentially follows the standard approach of
Refs.~\cite{Dienes:2021woi,Dienes:2022zgd,Dienes:2023ziv}.  However, in order to discuss the 
emergence of a hidden universal $n_{\rm eff}=3/2$ stasis from this example, it will prove useful 
to develop a different way of thinking about the dynamics of this system.   In particular, we shall 
seek to exploit one of the important symmetries that characterize this stasis.

\begin{figure}[t!]
\centering
\includegraphics[width=0.50\textwidth, height=0.38 \textwidth]{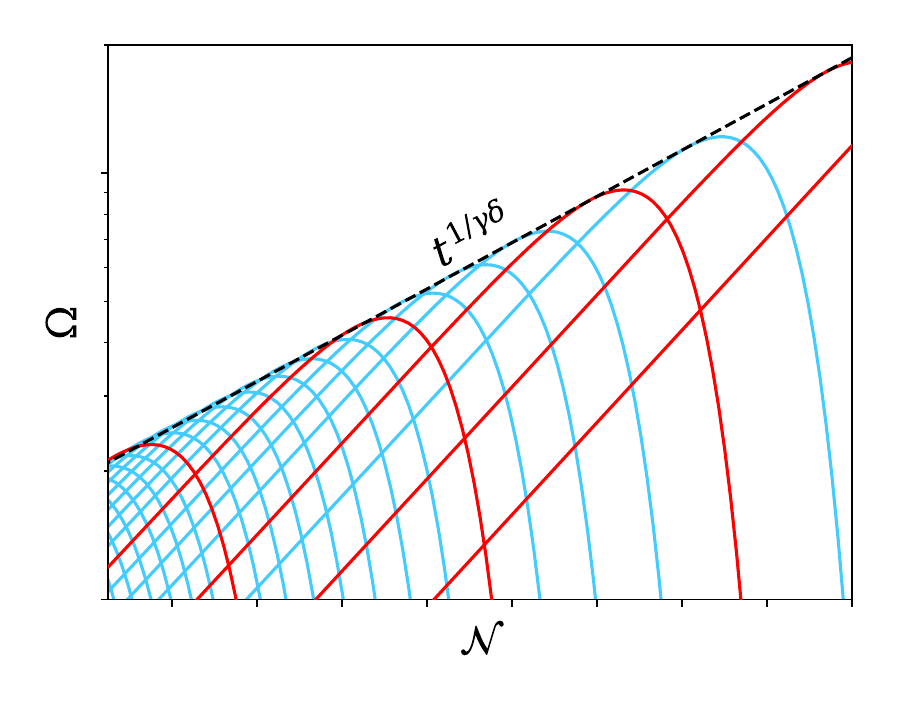}
\caption{A close-up of the stasis portion of Fig.~2 of Ref.~\cite{Dienes:2021woi}, showing the 
time-evolution of the individual matter abundances $\Omega_\ell(t)$ (blue and red curves) associated 
with a tower of states during matter/radiation stasis.  Abundances with smaller values of $\ell$ reach 
their maxima only after those with larger values of $\ell$ have already done so.   Also shown is the 
envelope function (dashed black line), which grows as $t^{1/\gamma\delta}$.  This figure illustrates 
the core ``self-similarity'' of our system when we are far from the ``edge effects'' associated with 
the top or bottom of the tower: as time progresses, the identity of the state with the largest abundance 
proceeds down the tower towards smaller values of $\ell$ (\ie, towards the right side of this figure) 
while the rest of this figure undergoes a common rescaling of both the horizontal and vertical directions.
Thus, as time progresses, we continually obtain increasingly shifted and rescaled versions of the original 
figure.  This self-similarity as time progresses is described mathematically in Eqs.~(\ref{self-similarity}) 
and~(\ref{self-similarity2}), and is a fundamental feature of stasis when stasis is realized through towers 
of states.  Those states whose abundances are equally spaced both horizontally and vertically (such as those 
plotted in red) have $\ell$-indices $\ell_n=\ell_0^n$ for a given $\ell_0$ and $n\geq 0$, $n\in \IZ$.
\label{fig:stasis}}
\end{figure}

To understand this symmetry, let us revisit the dynamics discussed above and consider how the individual 
abundances $\Omega_\ell(t)$ behave during a period of stasis in which their total 
$\Omega_M\equiv \sum_\ell \Omega_\ell(t)$ remains constant. This behavior is shown in 
Fig.~\ref{fig:stasis}, which is essentially a ``close-up'' of the stasis portion of Fig.~2 of 
Ref.~\cite{Dienes:2021woi}.  Despite the discrete nature of the individual states in our tower 
(and therefore of the curves shown in Fig.~\ref{fig:stasis}), for convenience we shall often consider 
the behavior of this system in the continuum limit in which the $\ell$-values of our states form a 
continuous variable, just as is done in Refs.~\cite{Dienes:2021woi,Dienes:2022zgd,Dienes:2023ziv}.
However, this will not be critical for our final results.

One way to study how this system evolves as a function of time is to focus on the behavior of each 
individual abundance $\Omega_\ell(t)$ during stasis.  Indeed, as we have seen in Eq.~(\ref{MGOmegal}), 
each abundance $\Omega_\ell(t)$ first rises before reaching a maximum value and then falling.  The 
sum of these abundances nevertheless remains constant, thereby producing a stasis epoch.  This 
approach was described in detail in the previous subsection.  Indeed, within this approach, the fact 
that the individual abundances $\Omega_\ell(t)$ sum to a constant is somewhat remarkable and perhaps 
even mysterious.

However, there is another way in which we might describe the time-evolution of this system --- one which 
describes the behavior of this system in a {\it collective}\/ manner, taking into account the behavior of 
the abundances corresponding to {\it all}\/ values of $\ell$ simultaneously and studying how they relate to 
each other as a function of time.  As we shall see, such an approach will have the distinct advantage of 
allowing us to understand {\it why}\/ stasis emerges in such systems, and how this stasis can be viewed 
as a collective phenomenon resulting from the underlying symmetries that these systems obey.

The primary characteristic of this collective behavior that will concern us here is the fact that the 
abundances within Fig.~\ref{fig:stasis} exhibit a {\it self-similarity}\/ across the tower as time 
evolves.  This self-similarity  is reflected in the fact that one cannot simply look at the abundances 
within Fig.~\ref{fig:stasis} and deduce what part of the tower is being shown.  Indeed, the spectrum of 
individual abundances $\lbrace \Omega_\ell\rbrace$ across our tower at any given time is simply a uniform 
rescaling of the spectrum of abundances at an earlier time, along with a simultaneous relabeling of their 
$\ell$-indices.  More specifically, each abundance $\Omega_\ell$ at a given time is a rescaled version 
of the value that a {\it different}\/ abundance $\Omega_{\ell'}$ had taken at an earlier time, where 
$\ell'$ is the identically rescaled value of $\ell$.   For example,  at any given moment,  the value 
of $\Omega_{100}$ is exactly twice what $\Omega_{200}$ had been a certain number of $e$-folds earlier.  
It is also exactly four times what $\Omega_{400}$ had been twice as many $e$-folds earlier, and so forth.

It is this property which leads to the self-similarity of Fig.~\ref{fig:stasis}.~  We can also express 
this self-similarity in terms of time-evolution in the {\it forward}\/ direction, without looking 
backward to earlier times.  Indeed, during a fixed interval of time-evolution, all abundances are  
magnified by a common factor which depends on the duration of the time interval.  However, these 
abundances are also each simultaneously relabeled in such a way that their new $\ell$-indices are 
divided by this same factor.  This procedure of simultaneously rescaling the abundances and inversely 
scaling their $\ell$-indices then yields a description of our system at a later time, and amounts to 
considering vertical time slices within Fig.~\ref{fig:stasis} that are increasingly to the right of 
the figure, corresponding to increasingly large values of $\calN$.  Thus, for example, if $\Omega_{400}$ 
reaches its maximum value at a given time, then at a certain later time it will be $\Omega_{300}$ 
that is reaching its maximum value, and this maximum value will be larger by a factor of $4/3$ than 
that previously reached by $\Omega_{400}$.  In all other respects, however, the basic pattern of 
interleaving abundances in this figure does not change.

Given this behavior, we see that time-evolution in this system boils down to two effects that operate 
simultaneously across the entire system:  a magnification of the individual abundances $\Omega_\ell$ 
along with a simultaneous shrinking of their indices $\ell$ (or more explicitly a relabeling of the 
indices across our system so that each abundance is now labelled with a smaller $\ell$-value).
In this sense we can therefore ascribe a time-evolution {\it not only to the abundances $\Omega$ 
but also to their $\ell$-indices}\/ as we pass increasingly toward the right in Fig.~\ref{fig:stasis}.  
Indeed, just as the abundances are changing with time, we may view their corresponding $\ell$-indices 
as changing in precisely the inverse way.

It may seem unusual to consider an index such as $\ell$ as having a time dependence.  However, this 
situation is somewhat analogous to that which arises in fluid dynamics, where one can either
concentrate on a particular ``packet'' of fluid as it proceeds along the fluid flow line, or instead 
concentrate on a fixed region of space and observe the different packets of fluid as they each pass 
through this region.  The former picture is analogous to concentrating on the time-evolution of a specific 
abundance --- an approach in which the index $\ell$ is unchanging --- while the latter picture is analogous 
to anchoring our attention to a fixed physical {\it condition}\/ (such as an abundance reaching its 
maximum value) and then observing different abundances sequentially satisfying this condition as the 
system evolves.  It is the continual change in the identity of the specific abundance satisfying this 
condition at any moment in time that is captured by the time-dependent index $\ell$.  Indeed, although 
we shall not do so in this paper, one might even go so far as to adopt a more general notation 
$\Omega_\ell(\calN)\to \Omega(\ell,\calN)$ to reflect the idea the physics of our system is actually 
described by an overall abundance function which depends on two variables, $\ell$ and $\calN$, the first 
variable specifying {\it which}\/ abundance we are considering and the second specifying {\it when}\/ that 
abundance is to be evaluated.  As time evolves (\ie, as $\calN$ increases), the overall abundance function 
$\Omega$ is magnified while the $\ell$ variable is inversely rescaled by the same factor.

We may also express this time-evolution mathematically.  As indicated in Fig.~\ref{fig:stasis} --- and 
as originally derived in Eq.~(3.27) of Ref.~\cite{Dienes:2021woi} --- this overall rescaling factor from 
any initial time $t_i$ to final time $t_f$ is given by $(t_f/t_i)^{1/\gamma\delta}$, or equivalently by 
$\exp\left( 3 \Delta \calN/\gamma \delta \barkappa\right)$ where the number of elapsed $e$-folds is 
given by $\Delta \calN = (\barkappa/3) \log(t_f/t_i)$.  Indeed, this magnification factor is nothing 
but the slope of the dashed envelope function in Fig.~\ref{fig:stasis}.  Time evolution in this system 
is therefore governed by the relation
\begin{equation}
\boxed{~~\Omega_\ell(\Ncal + \Delta \Ncal) ~=~ u\,\Omega_{u\cdot \ell}(\Ncal)~~}
\label{self-similarity}
\end{equation}
where
\beq
  u ~\equiv ~ \exp\left(\frac{3 \Delta \calN}{\gamma\delta\,\barkappa }
    \right)~=~ \left( \frac{t+\Delta t}{t}\right)^{1/\gamma\delta}~
  \label{ucorrected}
\eeq
and where $u\cdot \ell$ in Eq.~(\ref{self-similarity}) denotes the product of $u$ and $\ell$.  This 
equation tells us that under time-evolution two rescalings occur:  our individual abundances are 
magnified by the factor $u$, so that $\Omega\to u\Omega$, and these abundances are also simultaneously 
relabeled so that $\ell\to u^{-1}\ell$.  No other fundamental characteristics of this system are altered.   
Of course, the result in Eq.~(\ref{self-similarity}) is written in a manner in which the transformation 
for $\Omega$ is an active one while the transformation for $\ell$ is implicitly passive.   Writing this 
relation in terms of purely active transformations, we  instead have
\beq
  \boxed{~~  \Omega_{u^{-1}\ell}(\calN+\Delta \calN) ~=~ u\, \Omega_\ell(\calN)~.~}
  \label{self-similarity2}
\eeq

Given that $u$ is not necessarily an integer, it follows that $u\ell$ and $u^{-1}\ell$ will not generally 
be integers.  The results in Eqs.~(\ref{self-similarity}) and (\ref{self-similarity2}) thus hold only in 
the limit that our discretum of tower states can be approximated as a continuum, with $\ell$ stretching 
effectively to infinity, signifying an extremely large tower of states.  However, as demonstrated in
Refs.~\cite{Dienes:2021woi, Dienes:2023ziv}, this is generally an excellent approximation, even for a 
large but finite discretum of states.

The self-similar rescaling in Eqs.~(\ref{self-similarity}) and (\ref{self-similarity2}) is the 
fundamental underlying feature that gives rise to stasis when stasis is realized through a tower of states.
Indeed, during stasis, we see that we may regard the process of time translation as being described by the 
operational replacements
\begin{eqnarray}
  {\cal T}(\Delta \calN):~~~~~\Ncal &\to& \Ncal + \Delta \Ncal \nonumber\\
    H &\to&u^{-\gamma\delta}\, H \nonumber\\
    \Omega&\to& u \Omega \nonumber\\
    \ell &\to& u^{-1} \ell
\label{shift_operations}
\end{eqnarray}
where $H$ is the Hubble parameter.  
Stasis is then nothing more than the statement that certain quantities such as the total matter abundance 
$\Omega_M$ are {\it invariant}\/ under the time-evolution operator $\calT(\Delta \calN)$.  Indeed, using 
the self-similarity property in Eq.~(\ref{self-similarity}), we can immediately verify that the total matter 
abundance $\Omega_M$ is constant during stasis:
\begin{eqnarray}
    \Omega_M(\Ncal + \Delta \Ncal) &=& \int d\ell \, \Omega_\ell (\Ncal + \Delta \Ncal) \nonumber\\
    &=& \int d\ell \, u \,\Omega_{\ell\cdot u} (\Ncal ) \nonumber\\
    &=& \int d\ell'\, \Omega_{\ell'} (\Ncal) ~~~{\rm where}~ \ell'=u\ell \nonumber\\
    &=& \Omega_M (\Ncal).
\label{totpreserved}
\end{eqnarray}

There is, of course, a natural limit to the extent to which this time-evolution can continue:
the ongoing rescaling process for the individual abundances and their $\ell$-indices  ultimately ends 
when the largest abundance approaches unity.  As described in Ref.~\cite{Dienes:2021woi}, this occurs 
when the index $\ell_\ast$ (indicating the state with the largest abundance at any given time) reaches 
the bottom of the tower.  This then signifies the end of the  stasis epoch.  Nevertheless, within the 
stasis epoch, the self-similar scaling property described in Eqs.~(\ref{self-similarity}) 
and~(\ref{self-similarity2}) remains a good symmetry and continues to ensure that the total abundance 
remains constant. 

As described above, we can also understand the time-evolution in our system by specifying a fixed 
``reference'' condition at each moment in time and studying how the identity of the state satisfying 
this condition changes as we time-evolve towards the right side of Fig.~\ref{fig:stasis}.  For example, 
at any time $\calN$ we can define $\ell_\ast$  as the $\ell$-value of that state within the tower which 
is instantaneously reaching its maximum abundance. Identifying this maximum-abundance time as 
$\tau_{\ell_\ast}= \Gamma_{\ell_\ast}^{-1}\sim (\ell_\ast)^{-\gamma \delta}$, we immediately find that 
$\calN$ and $\ell_\ast$ are related via 
\beq
  \calN ~=~ C - \frac{\gamma \delta \barkappa }{3} \, \log \,\ell_\ast
\label{lastdef}
\eeq
where $C$ is a time-independent constant.  This establishes a firm relationship between $\calN$ and 
$\ell_\ast$ and demonstrates that $\ell_\ast$ decreases as $\calN$ increases, with
\beq
  \partial_\calN \log\,\ell_\ast~=~ - \frac{3}{\gamma\delta\barkappa}~.
\eeq
Indeed, the minus sign in this relation tells us that the identity of the state just reaching its maximum 
abundance at time $\calN$ is continually shifting down the tower as time increases.

Of course, this result would continue to apply if $\ell_\ast$ were instead identified as any other 
specific reference $\ell$-value at any given time.  We shall therefore more generally simply write
\beq
  \partial_\calN \log\,\ell  ~=~ -\frac{3}{\gamma\delta\barkappa}~,
\label{logell}
\eeq
which again reflects our view of any particular $\ell$-index as continually evolving with time due to 
the continuing relabeling of our abundances.  Indeed, this result is consistent with the continual 
shifting of $\ell$-indices indicated within $\calT(\Delta \calN)$ in Eq.~(\ref{shift_operations}).

Given that the passage of time corresponds to specific $\ell$-values such as $\ell_\ast$ passing to 
smaller and smaller values down the tower,  it is natural to define the logarithmic {\it velocity}\/ 
per unit $e$-fold with which this happens:
\beq
   v ~\equiv~ \partial_\calN 
   \log\, \ell_\ast ~=~ -\frac{3}{\gamma\delta \barkappa}~,
\label{shiftvelocity}
\eeq
or equivalently
\beq
   v ~\equiv~ \partial_\calN 
   \log\, \ell  ~=~ -\frac{3}{\gamma\delta \barkappa}~.
\label{shiftvelocity2}
\eeq
We thus have $u=\exp(v\Delta \calN)$, or equivalently 
\beq
   v~=~ \frac{1}{\Delta \calN} \log u~.
\label{vu} 
\eeq
Indeed, the velocity $v$ is constant during stasis, indicating that $\log \,\ell$ changes at a fixed 
rate per unit $e$-fold.

\subsection{Obtaining $n_{\rm eff}=3/2$ from the tower \label{sec:n32tower}}

Given this self-similarity symmetry, we now turn to the value of $n$ that appears in Eq.~(\ref{genn}).  
Our assertion is that all of the different stases that have been examined in the literature --- despite 
the different appearances of their corresponding pumps --- actually have a common effective value 
$n_{\rm eff}=3/2$ during stasis, where $n_{\rm eff}$ is defined in Eq.~(\ref{ndef2}).  We shall now 
demonstrate that this is true for the tower-based matter/radiation stasis we have been discussing above.
Indeed, the definition of $n_{\rm eff}$ given in Eq.~(\ref{ndef2}) allows $n_{\rm eff}$ to take 
non-integer values and most importantly to {\it evolve}\/ dynamically as a function of time as our 
system approaches stasis.

Motivated by the form for $n_{\rm eff}$ in Eq.~(\ref{ndef2}), we begin by evaluating the behaviors 
of $P_{M\gamma}^{(\rho)}$ and $\rho_M$ under time-evolution.  Continuing onward from 
Eq.~(\ref{self-similarity}), we can first evaluate the time-evolution of the abundance pump 
$P_{M\gamma}$ during stasis, where
\beq
      P_{M\gamma}(\calN)~\equiv~
      \sum_\ell\, \Gamma_\ell\, \Omega_\ell(\calN)~
      ~\approx~
    \int d\ell\, \Gamma_\ell \,
     \Omega_\ell(\calN)~
\label{abpump}
\eeq
and where we have passed to the continuum limit.  However, our assumed scaling behavior for the decay 
widths $\Gamma_\ell$ across the tower indicates that
\beq
\Gamma_\ell~\sim~  
\left(m_\ell\right)^\gamma 
 ~\sim~ \ell^{\gamma\delta}~.
\eeq
Use of Eq.~(\ref{self-similarity}) then immediately tells us that
\beqn
P_{M\gamma}(\calN+\Delta \calN) &\sim &
  \int d\ell \, \Gamma_\ell \,
        \Omega_\ell(\calN+ \Delta \calN) \nonumber\\
        &=&
        \int d\ell \, \Gamma_\ell \,
        u\, \Omega_{u\cdot \ell}(\calN) \nonumber\\
&=&
u^{-\gamma \delta} \int d\ell' 
  \, \Gamma_{\ell'} \, \Omega_{\ell'}(\calN) ~~{\rm where}~ \ell'= u\ell\nonumber\\
&=& 
u^{-\gamma \delta} \,P_{M\gamma}(\calN)~
\label{Pscaling}
\eeqn
or equivalently 
\beq  
         P_{M\gamma}(t+\Delta t) ~=~
         \left(
         \frac{t}{t+\Delta t}\right)\,P_{M\gamma}(t)~
\eeq 
where $\Delta t$ is the elapsed time.   
In the $\Delta t\gg t$ limit,
we thus see that our pump scales inversely with the elapsed time, as required, 
and yields the result
\beq 
\partial_\calN P_{M\gamma}(\calN) 
~=~ -\frac{3}{\barkappa}\,
P_{M\gamma}(\calN)~.
\label{startderiv}
\eeq
However, $P_{M\gamma}^{(\rho)} \sim H^2 P_{M\gamma}$.
We thus have
\beqn
 \partial_\calN P_{M\gamma}^{(\rho)}
 ~&=&~ -\frac{6}{\barkappa} \,P_{M\gamma}^{(\rho)}
     -\frac{3}{\barkappa} \,P_{M\gamma}^{(\rho)}\nonumber\\
 ~&=&~ -\frac{9}{\barkappa} \,P_{M\gamma}^{(\rho)}~.
\label{numer}
\eeqn

Likewise we know that $\rho_M\sim H^2 \Omega_M$.
During stasis, this in turn implies that
\beq
 \partial_\calN \rho_M
 ~=~ -\frac{6}{\barkappa}\, \rho_M~.
\label{denom}
\eeq
Thus, combining Eqs.~(\ref{numer}) and (\ref{denom}), 
we have
\beq
n_{\rm eff} ~\equiv~ 
\frac{\partial_\Ncal \log P_{M\gamma}^{(\rho)}}{\partial_\Ncal \log \rho_M}~
=~ \frac{9/\barkappa}{6/\barkappa}~=~\frac{3}{2}~
\label{endderiv} 
\eeq
during stasis.

Interestingly, we observe that our result $n_{\rm eff}=3/2$ does not actually require knowing the 
specific form of the pump $P$.  Indeed, all that is required in order to obtain $n_{\rm eff}=3/2$ is 
that our abundance pump satisfy Eq.~(\ref{Pscaling}), or equivalently that $P(t)\sim 1/t$ where $t$ 
is the elapsed time.   We have already seen this behavior in Sect.~\ref{relationtopumptimescaling}.~
However, despite the straightforward nature of this proof, this result is extremely non-trivial.
In particular, it is remarkable that the self-similar scaling behavior in Eq.~(\ref{self-similarity})
manages to simultaneously yield not only a constant total abundance via Eq.~(\ref{totpreserved}) but 
also a $1/t$ scaling for the pump via Eq.~(\ref{Pscaling}).

{\it A priori}\/, one might have suspected that $n_{\rm eff}$ must always be an integer.  
Certainly the form of our pump --- linear in energy densities --- suggests that $n=1$.  Moreover, 
in simple cases in which $Z$ is a constant in Eq.~(\ref{genn}), we would indeed have $n_{\rm eff}=n= 1$.  
There must therefore be an effect introduces a time-dependence into $Z$, thereby {\it deforming}\/ the 
value of $n_{\rm eff}$ away from such an integer value and leaving us with the universal result 
$n_{\rm eff}=3/2$.  Fortunately, it is not difficult to determine what gives rise to this deformation: 
it is the existence of the tower itself and the fact that our self-similarity mapping involves continual 
shifts down the tower as time advances.   

To understand this mathematically, we can derive an alternative expression for $n_{\rm eff}$ which makes 
this clear.  Starting from the abundance pump $P_{M\gamma}(\calN)$ in Eq.~(\ref{abpump}) we immediately 
have the energy-density pump
\beq
    P_{M\gamma}^{(\rho)}~=~  \int d\ell\,
    \Gamma_{\ell} \,\rho_{\ell}~.
\label{Prho}
\eeq
As written, this pump has three components:  the measure $d\ell$, the decay width $\Gamma_\ell$, and 
the energy density $\rho_\ell$.  Of course, strictly speaking, only one of these quantities --- the 
energy density $\rho_\ell$ --- is explicitly time-dependent (or $\calN$-dependent).  Indeed, $\ell$ is 
nothing more than a dummy variable within this integral.  However, we have already seen that a shift in 
time can be viewed as inducing induces changes not only in our abundances but also in their effective 
$\ell$-values, and this in turn will induce a change in the decay widths since these widths likewise 
depend on $\ell$.  Thus, it is legitimate to treat all three of these quantities as $\calN$-dependent.   

With this understanding, it follows from Eq.~(\ref{Prho}) that
we can generally write the time variation of our energy-density pump as
\beq
 \partial_\calN \log\, P_{M\gamma}^{(\rho)}
 ~=~ \partial_\calN \log \,{\ell }
 \,+\, \partial_\calN \log \,\Gamma_{\ell }
 \,+\, \partial_\calN \log \, \rho_{\ell }~~~
\label{threeterms}
\eeq
where the three terms on the right side of Eq.~(\ref{threeterms}) result respectively from the 
time-variations of the measure $d{\ell }$, the decay width $\Gamma_{\ell }$, and the energy density
$\rho_{\ell }$ within Eq.~(\ref{Prho}).  However, we now recall the similar expression
for $\rho_M$, specifically
\beq
\rho_M ~=~ \int d{\ell } \,\rho_{\ell }~,
\eeq
from which it likewise follows that
\beq
 \partial_\calN \log\, \rho_M
 ~=~ \partial_\calN \log \,{\ell }
 \,+\, \partial_\calN \log \, \rho_{\ell }~.~~
\label{twoterms}
\eeq
This relation allows us to bundle two of the terms on the right side of Eq.~(\ref{threeterms}) into a 
single term, yielding
\beq
 \partial_\calN \log\, P_{M\gamma}^{(\rho)} ~=~
  \partial_\calN \log\, \rho_M \,+\, \partial_\calN \log \,\Gamma_{\ell }~,
\eeq
or equivalently
\beqn
n_{\rm eff} \,\equiv\,
\frac{\partial_\Ncal \log P_{M\gamma}^{(\rho)}}{\partial_\Ncal \log \rho_M}
\,&=&\, 1+ \frac{\partial_\calN \log \,\Gamma_{\ell }}{\partial_\calN \log \,\rho_M} \nonumber\\
\,&=&\, 1+ \gamma\delta\,\frac{\partial_\calN \log \,{\ell }}{\partial_\calN \log \,\rho_M}  ~~~~~~~
\label{altexpression}
\eeqn
during stasis.

The final expression in Eq.~(\ref{altexpression}) exposes the origin of the deformation of $n_{\rm eff}$:  
this deformation away from the expected value $n=1$ occurs as the result of the continual shifting of 
our states down the tower, with $\ell $ decreasing as $\calN$ increases.  It is also straightforward 
to calculate the magnitude of this deformation.  Indeed, given the results in Eqs.~(\ref{lastdef}) 
and (\ref{denom}) we have
\beq
\frac{\partial_\calN \log \,{\ell }}{\partial_\calN \log \,\rho_M} ~=~ 
    \frac{-3/\gamma\delta\barkappa}{-6/\barkappa}  ~=~
    \frac{1}{2\gamma\delta}~.
\eeq
We thus find 
\beq
    n_{\rm eff} ~=~ 1 + \gamma\delta\,
   \,\left(\frac{1}{2\gamma\delta}\right) =~ \frac{3}{2}~,
\eeq
precisely as before.

We can also demonstrate how our system {\it evolves toward}\/ $n_{\rm eff}=3/2$ regardless of the 
initial conditions.  This feature is ultimately guaranteed by the fact that the stasis state is a global 
attractor, but it is instructive to see precisely how the value of $n_{\rm eff}$ evolves as a result of this 
attractor behavior.

Towards this end, let us revisit the attractor analysis of this system that was presented in 
Ref.~\cite{Dienes:2023ziv}, only now adding a study of $n_{\rm eff}$ to the analysis.  In 
Ref.~\cite{Dienes:2023ziv}, the possible trajectories of this system were studied within the 
$(\Omega_M,\calH)$ plane, where $\calH(t)$ is the ratio of the Hubble parameter $H(t)$ at any given 
time $t$ to the value which it would have at that time if the universe were in stasis:
\begin{equation}
  \mathcal{H}(t) ~\equiv~ H(t)\left(\frac{3t}{\barkappa}\right)~.
\end{equation}
Here $\barkappa$ is the stasis value of $\kappa$, where $\kappa$ is defined in Eqs.~(\ref{kappadef}) 
and~(\ref{kappadef2}); for a universe consisting of only matter and radiation we have
\beq
 \kappa ~=~ \frac{6}{4-\Omega_M}~,
\eeq
with $\barkappa$ similarly related to $\barOmega_M$.
 Using these variables, we can then write our differential equations for $\Omega_M$ and $\calH$ in the 
 relatively simple forms~\cite{Dienes:2023ziv}
\beqn
\partial_\calN \Omega_M &=&
\Omega_M \left[ 
   (1-\Omega_M) - \frac{1}{\calH} (1-\barOmega_M)\right] \nonumber\\
\partial_\calN \calH &=& 
 \left(\frac{4-\barOmega_M}{2}\right)
    -\calH 
 \left(   \frac{4-\Omega_M}{2}\right) ~
\label{diffeqsmatterradiation}
\eeqn 
from which we immediately verify that both derivatives are zero only for the stasis solution 
$\Omega_M=\barOmega_M$ and $\calH=1$.  Indeed, for any initial values of $\Omega_M$ and $\calH$, our 
system evolves along a specific trajectory within the $(\Omega_M,\calH)$ plane which is determined by 
these differential equations.  These trajectories are plotted in the left panel of Fig.~2 of 
Ref.~\cite{Dienes:2023ziv} for the case in which $\barOmega_M=1/2$, and one finds that these 
different possible trajectories always lead to the stasis solution.  Note that these trajectories are 
independent of where our system started --- in other words, these trajectories do not carry any ``prior'' 
knowledge about the initial times at which the fields involved in our stasis were originally produced.   
This is ultimately guaranteed by the fact that the dynamical equations for this system are autonomous.

\begin{figure}
\centering 
 \includegraphics[keepaspectratio, width=0.53\textwidth] {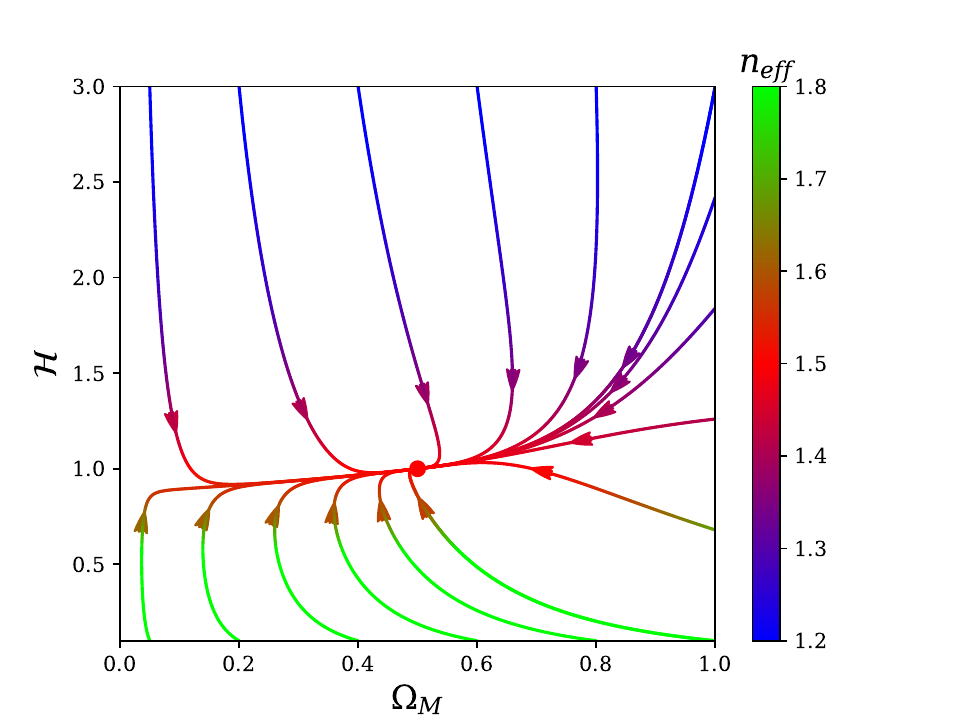}
\caption{
Attractor plot for a matter/radiation stasis with $\alpha\delta=1$ and $\gamma\delta=7$, illustrating 
that all trajectories that our system can follow within the $(\Omega_M,\calH)$-plane lead to the stasis 
solution $(\barOmega_M,\calH)= (1/2,1)$.   However, unlike the left panel of Fig.~2 of 
Ref.~\cite{Dienes:2023ziv}, we have now plotted these trajectories with colors indicating the 
corresponding values of $n_{\rm eff}$ that are realized at each moment in the evolution of our system.  
We observe that $n_{\rm eff}\not= 3/2$ for generic points during the evolution of our system, but that 
$\neff\to 3/2$ as the fixed-point solution is approached.
\label{fig:attractorTowerNeff}}
\end{figure}

The fact that all of these trajectories lead to the stasis solution verifies that our stasis solution 
is a global attractor.  However, let us now track the manner in which $n_{\rm eff}$ evolves as our system 
travels along these trajectories.  In general, $n_{\rm eff}$ for this system is given by
\beq
  n_{\rm eff} ~=~  \frac{3}{2} - \frac{1}{2}\frac{\partial_\Ncal \ln\Omega_M}{\partial_\Ncal \ln \rho_M} 
    - \frac{\partial_\Ncal \ln\Hcal}{\partial_\Ncal \ln \rho_M}~.
  \label{nefftrajectory}
\eeq
This result does not assume stasis, and thus allows us to evaluate $n_{\rm eff}$ at any moment during the 
evolution of our system.   Our results are shown in Fig.~\ref{fig:attractorTowerNeff}, where we have shown 
the same trajectories as in the left panel of Fig.~2 of Ref.~\cite{Dienes:2023ziv} but where we have now 
introduced  {\it colors}\/ along these trajectories in order to indicate the corresponding values of 
$n_{\rm eff}$.  As evident from this figure, our system experiences many different values of $n_{\rm eff}$ 
as it travels along a given trajectory.  Indeed, although difficult to see in this figure, there even exist 
trajectories within the $(\Omega_M,\calH)$ plane along which the evolution of $n_{\rm eff}$ is non-monotonic.
However, in all cases we find that $n_{\rm eff}\to 3/2$ as the system approaches the stasis solution.  
Indeed, this is already apparent from Eq.~(\ref{nefftrajectory}), given that the stasis state is one in 
which the time-derivatives of $\Omega_M$ and $\calH$ appearing in this equation both vanish.

Despite appearances, it is important to note that $n_{\rm eff}$ is not a simple  or direct function of 
the coordinates $(\Omega_M,\calH)$ of our system within the $(\Omega_M,\calH)$ plane.  Instead, we see 
from Eq.~(\ref{nefftrajectory}) that $n_{\rm eff}$ is also sensitive to the instantaneous {\it slope}\/ 
of the trajectory within this plane, and even the {\it curvature}\/ of this trajectory, since these 
quantities govern the relative rates at which $\Omega_M$ and $\calH$ vary.   Thus $n_{\rm eff}$ captures 
a considerable amount of information about our system beyond its mere coordinates  within the 
$(\Omega_M,\calH)$-plane.  That said, since each point in the $(\Omega_M,\calH)$ plane lies along a 
single trajectory, knowledge of the location of our system within this plane is ultimately sufficient 
to determine $n_{\rm eff}$, assuming that full knowledge of the dynamics of our system is available to us.
Thus the value of $n_{\rm eff}$ is indeed determined uniquely for each location, but in a complex way 
that also depends on the shape of the trajectory passing through that point.

\subsection{Other tower-based stases:  Vacuum-energy/matter stasis and beyond}

Our discussion thus far in this section has focused on matter/radiation stasis in which the transition 
from matter energy density to radiation energy density occurs through particle decays --- \ie, through a 
pump of the form in Eq.~(\ref{abpump}).  However, there also exists another tower-based stasis which has 
been examined in the prior literature~\cite{Dienes:2023ziv} and which employs an entirely different sort 
of $n=1$ pump:  this is a vacuum-energy/matter stasis which involves a tower of scalar fields whose zero 
modes experience a {\it phase transition}\/ from an overdamped phase (during which the corresponding 
energy density is associated with vacuum energy) to an underdamped phase (during which it is associated 
with matter).   This underdamping transition thus effectively furnishes us with a pump which converts 
vacuum energy to matter --- indeed, one whose effects are ultimately counterbalanced by the effects of 
cosmological expansion, thereby leading to a vacuum-energy/matter stasis.

Although a fully dynamical study of this system was presented in Ref.~\cite{Dienes:2024wnu}, we shall 
here adopt the simplifying assumptions that were adopted in Ref.~\cite{Dienes:2023ziv}.  In particular, 
for each field $\phi_\ell$ in the tower, this transition is taken to occur fully and instantaneously at 
the time $t_\ell$ at which the critical underdamping condition $3H(t_\ell)=2 m_\ell$ is satisfied.  Prior 
to the time $t_\ell$, the energy associated with the $\phi_\ell$ is to be considered vacuum energy, 
while that after $t_\ell$ is to be considered matter.  Moreover, prior to $t_\ell$, we shall treat the 
energy density associated with $\phi_\ell$ as a fluid whose equation-of-state parameter $w_\Lambda$ is 
a constant close to but slightly greater than $-1$ (so as to avoid certain singularities that arise 
when $w_\Lambda= -1$ but which are irrelevant for our discussion).  Likewise, the matter phase for each 
$\phi_\ell$ field after $t_\ell$ is modeled as a fluid with constant $w=0$.

Under these assumptions it is straightforward to determine the mathematical form of the pump for this 
system.  With $\rho_\ell$ now representing the vacuum-energy density associated with $\phi_\ell$, we 
find that 
\beq
  \rho_\ell(t) ~=~ \rho_\ell^{(0)} \, a^{-3(1+w_\Lambda)} \,\Theta(t_\ell-t)~
\eeq
where $\rho_\ell^{(0)}$ is the value of $\rho_\ell(t)$ at the initial production time $t^{(0)}$.
We thus find
\beq
\frac{d\rho_\ell}{dt} ~=~
   -3(1+w_\Lambda) H \rho_\ell -
   \rho_\ell \, \delta(t_\ell-t)~
\label{justbelowhere}
\eeq
whereupon
\beq
\frac{d\rho_\Lambda}{dt} ~=~ 
 \sum_\ell \frac{d\rho_\ell}{dt} ~=~
   -3(1+w_\Lambda) H \rho_\Lambda -
   \sum_\ell \rho_\ell \, \delta(t_\ell-t)~.
\label{justbelowhere2}
\eeq
We are therefore faced with evaluating the sum in the final term.  In order to evaluate this sum, we 
pass to a continuum limit in which we replace the discrete spectrum of underdamping times $t_\ell$ with 
a continuous variable $\hat t$.  We can likewise view the discrete spectrum of energy densities 
$\rho_\ell$ and corresponding abundances $\Omega_\ell$ as continuous functions 
$\widetilde \rho({\hat t})$ and $\widetilde \Omega(\hat t)$, where the states are now indexed by 
the continuous ${\hat t}$-variable corresponding to their underdamping times.  Thus, for example, 
$\widetilde \rho(\hat t)$ is the differential energy density associated  with those states in the 
tower which are decaying precisely at the time $\hat t$, and $\widetilde \Omega(\hat t)$ is similarly 
related to the abundance of those states in the tower which are decaying precisely at the time $\hat t$.
The $\ell$-sum in  Eq.~(\ref{justbelowhere2}) then becomes an integral over $\hat t$, \ie, 
$\sum_\ell \to \int d{\hat t}\, n_{\hat t}(\hat t)$, where $n_{\hat t}(\hat t)$ is the density of 
states per unit ${\hat t}$, evaluated at the location within the tower (\ie, for the value of $\ell$) 
for which $t_\ell=t$:
\beq
  n_{\hat t}(t) ~\equiv~ \biggl|\frac{d\ell}{d t_\ell} \biggr|_{t_\ell=t} ~.
  \label{densityofstates}
\eeq
We can thus evaluate the final term in
Eq.~(\ref{justbelowhere2}), obtaining
\beqn
 \sum_\ell \rho_\ell\, \delta(t_\ell-t) 
 ~&\to&~ 
 \int d \hat t \, n_{\hat t} (\hat t) \,
    \widetilde \rho(\hat t) \,\delta(\hat t-t) \nonumber\\
    &=& ~n_{\hat t}(t)\, \widetilde \rho(t)~.
\label{intermed}
\eeqn 
Of course, this passage from the $\ell$-sum to the $\hat t$-integral involves a number of approximations 
which are discussed in Refs.~\cite{Dienes:2021woi,Dienes:2023ziv}, but these play no essential role in 
the following for times which are near neither the initial production time nor the underdamping time 
of the lightest tower constituent.  Inserting Eq.~(\ref{intermed}) into Eq.~(\ref{justbelowhere2}), 
we thus obtain
\beq
  \frac{d\rho_\Lambda}{dt} ~=~ - 3(1+w_\Lambda )H\rho_\Lambda - n_{\hat t}(t) \, \widetilde\rho(t)~, 
  \label{eom1}
\eeq
from which we can identify the pump terms 
\beqn
  P_{\Lambda M}^{(\rho)}(t) &=&
    n_{\hat t}(t)\, \widetilde\rho(t) ~\nonumber\\
    \Longrightarrow ~~ P_{\Lambda M}(t) & = &
    n_{\hat t}(t) \,\widetilde\Omega(t)~.~~~~~~~
  \label{phasetransitionpump}
\eeqn
As predicted, these pumps are linear in the corresponding energy densities or abundances, thereby 
furnishing us with additional examples of apparent $n=1$ pumps.  These phase-transition pumps also 
make intuitive sense, telling us that the instantaneous rate of energy transfer from vacuum energy to 
matter  at any moment in this system is simply given by the amount of energy associated with those 
fields which happen to be undergoing the underdamping phase transition precisely at that moment, 
suitably multiplied by the density of states within that part of the tower.

\begin{figure}
\centering
\includegraphics[width=0.45\textwidth, height=0.38 \textwidth]{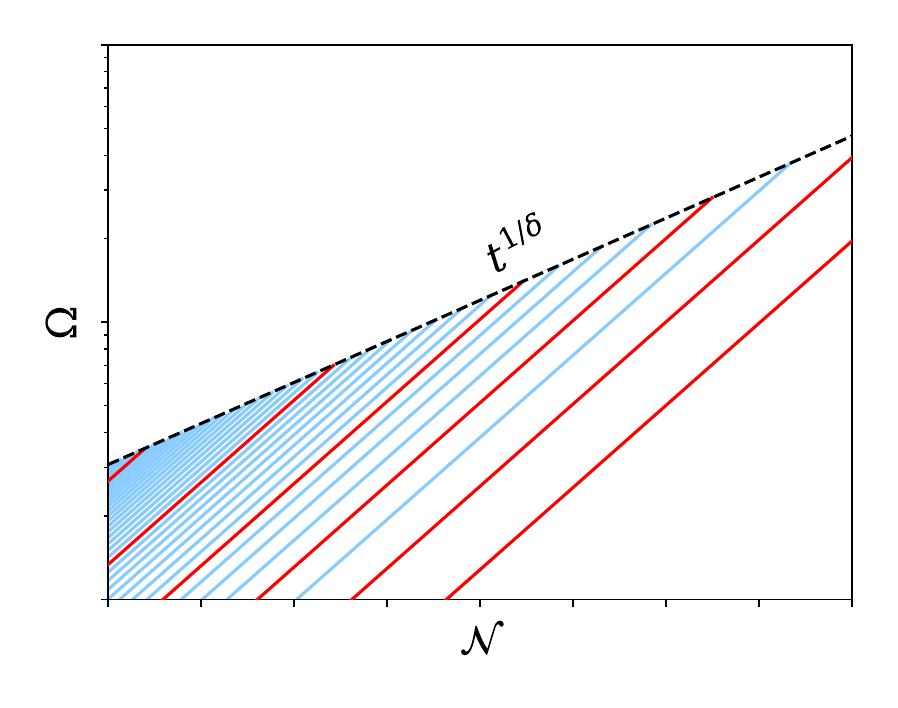}
\caption{Same as Fig.~\ref{fig:stasis}, but now for a vacuum-energy/matter stasis utilizing an $n=1$ pump based not on particle decay, but rather on an underdamping transition, as described in the text. 
Each line in this figure (either red or blue) corresponds to the vacuum-energy abundance $\Omega_\ell$ of a different state within the tower during stasis, first rising linearly on such a log plot during stasis before suddenly converting to matter when $3H(t_\ell)=2m_\ell$ and thereby no longer making contributions to vacuum energy.
The ``envelope'' function for these abundances (dashed black line) scales as $\sim t^{1/\delta}$.
Just as with in Fig.~\ref{fig:stasis},
this figure exhibits a `self-similarity'':  as time progresses, the identity of the state with the largest abundance proceeds down the tower towards smaller values of $\ell$ (\ie, towards the right side of this figure) while the rest of this figure undergoes a common rescaling of 
both the horizontal and vertical directions but is otherwise the same.
This self-similarity as time progresses is described mathematically in Eqs.~(\ref{self-similarity}) and
(\ref{self-similarity2}), but with a modified value of the rescaling factor $u$, as described in the text.
}
\label{fig:stasis3}
\end{figure}

The rest of our analysis proceeds exactly as for the matter/radiation stasis based on particle decay, 
as described in Sect.~\ref{sect:towerbasedreview}.~  In particular, under the same assumptions for the 
initial abundances $\Omega_\ell^{(0)}$ and mass spectra $m_\ell$ as in Eqs.~(\ref{scalings1}) 
and~(\ref{scalings2}), we find that each vacuum-energy abundance $\Omega_\ell(t)$ in this system rises 
linearly (on a log plot) from the initial production time until the time at which it experiences the 
underdamping phase transition.  After this time the abundance no longer contributes to the vacuum energy 
and thus effectively ``disappears'' from any tally of vacuum energy.  This behavior is shown in 
Fig.~\ref{fig:stasis3}, which is the analogue of what we previously found in Fig.~\ref{fig:stasis} 
for the matter/radiation system described in Sect.~\ref{sect:towerbasedreview}.  Indeed, 
$\widetilde \Omega(\calN)$ is nothing other than the dashed black envelope function in 
Fig.~\ref{fig:stasis3}.

As we see, Fig.~\ref{fig:stasis3} exhibits the same sort of self-similarity as shown in 
Fig.~\ref{fig:stasis}, with time-evolution corresponding to a process under which the individual 
vacuum-energy abundances $\Omega_\ell$  are rescaled and simultaneously relabeled with new $\ell$-indices.
Indeed, these abundances also satisfy the fundamental self-similarity equations in 
Eqs.~(\ref{self-similarity}) and (\ref{self-similarity2}).  The chief difference, however, is the form 
of the envelope function: rather than scale as $t^{1/(\gamma \delta)}$, as in Fig.~\ref{fig:stasis}, it 
now instead scales as $t^{1/\delta}$.  This, of course, makes sense, since there are no decays in this 
system and therefore $\gamma$ is no longer a relevant parameter.  This change in the envelope function 
then implies a change in the corresponding rescaling factor $u$.  Indeed, the vacuum-energy abundances 
$\Omega_\ell(t)$ in this system continue to obey Eqs.~(\ref{self-similarity}) and (\ref{self-similarity2}), 
only now with 
\beq
u ~\equiv ~ \exp\left(\frac{3 \Delta \calN}{\delta\,\barkappa }
  \right)~=~ \left( \frac{t+\Delta t}{t}\right)^{1/\delta}~.
\label{ucorrected2}
\eeq

Given this, we see that the time-evolution operator $\calT(\Delta\calN)$ continues to correspond to the 
shifts in Eq.~(\ref{shift_operations}), only with this new value of  $u$.  We also find, precisely as 
in Eq.~(\ref{totpreserved}), that the total vacuum-energy abundance 
$\Omega_\Lambda(t) \equiv \sum_\ell \Omega_\ell(t)$ is a constant.  This then confirms that this system 
also experiences a stasis --- this time a vacuum-energy/matter stasis.

We can also verify that this self-similarity symmetry ensures the proper behavior of our 
vacuum-energy/matter abundance pump $P$.  Given the form of the pump in Eq.~(\ref{phasetransitionpump}), 
we have
\beq
 P_{\Lambda M}(\calN+\Delta \calN) ~=~
   n_{\hat t} (\calN+\Delta \calN) \, \widetilde \Omega(\calN+ \Delta \calN)~,
\eeq
where we recall that $n_{\hat t}(\calN)$ and $\widetilde\Omega(\calN)$ respectively denote the density of 
states and the  vacuum-energy abundance for that part of the tower corresponding to the value of $\ell$ 
for which $t_\ell=t$, where $t$ is the time corresponding to the number of $e$-folds $\calN$.
Given this, it is straightforward to verify that if $\ell$ has a value such that $t_\ell=t$, then the new 
condition $t_{\ell'}=t+\Delta t$ is solved by $\ell'= u^{-1} \ell$, consistent with 
Eq.~(\ref{shift_operations}).  In other words, we have
\beq
  \widetilde \Omega(\calN+\Delta \calN) ~=~ u \,\widetilde\Omega(\calN)~,
\eeq
which is consistent with the idea that $\widetilde \Omega(\calN)$ is essentially the envelope function in Fig.~\ref{fig:stasis3}.~
Likewise, under time-evolution we also have 
\beq
  n_{\hat t}(\calN+\Delta\calN) ~=~
    \left|\frac{\partial(u^{-1}\ell)}{\partial (u^\delta t)}
    \right| ~=~ u^{-1-\delta}\, n_{\hat t}(\calN)~.
\eeq
Putting these results together then yields
\beqn
  P_{\Lambda M}(\calN+\Delta \calN) ~&=&~ u^{-1-\delta} \,n_{\hat t}(\calN)
    \,u\,\widetilde \Omega(\calN) ~~~~~\nonumber\\
  &=&~ u^{-\delta} \,P_{\Lambda M}(\calN)~.
\eeqn
This confirms that our abundance pump $P$ in Eq.~(\ref{phasetransitionpump}) indeed scales as $1/t$ for 
large $t$, as required.  The rest of our analysis then proceeds exactly as before, only with $\gamma=1$.

The fact that we achieve stasis with a pump scaling as $1/t$ implies that $n_{\rm eff}=3/2$ during stasis.  
Once again, an explicit demonstration of this proceeds along the lines of Eqs.~(\ref{startderiv}) 
through (\ref{endderiv}).  Moreover, as demonstrated in Ref.~\cite{Dienes:2023ziv}, this stasis is also 
a global attractor.

We can also examine how our value of $n_{\rm eff}$ {\it approaches}\/ 3/2 as our system travels along 
a dynamical trajectory within the $(\Omega_\Lambda,\calH)$ plane.  In general, these trajectories are 
governed by the differential equations~\cite{Dienes:2023ziv}
\beqn
  \partial_\calN \Omega_\Lambda &=& -3 w_\Lambda \Omega_\Lambda
    \left(1-\Omega_\Lambda\right) - \frac{3}{2}\,\eta \,\Omega_\Lambda 
    \left(1+w_\Lambda \Omega_\Lambda\right) \nonumber\\
  \partial_\calN 
    \calH &=& \frac{3}{2}(1 + w_\Lambda \barOmega_\Lambda) 
    - \frac{3}{2}\calH (1 + w_\Lambda \Omega_\Lambda) ~.
  \label{dyn} 
\eeqn
Although these equations for our vacuum-energy/matter stasis superficially resemble their analogues
for matter/radiation stasis in Eq.~(\ref{diffeqsmatterradiation}), there is one crucial difference:
when these equations are expressed in terms of derivatives with respect to $\calN$ (as is done here) 
rather than with respect to the time $t$ (as in Ref.~\cite{Dienes:2023ziv}), the first of these equations 
loses all dependence on $\calH$.  Thus, even though the time-evolution of $\calH$ depends on both 
$\Omega_\Lambda$ and $\calH$ itself, we see that $\Omega_\Lambda$ evolves according to the single-variable 
top equation within Eq.~(\ref{dyn}) without sensitivity to $\calH$.  It turns out that this 
``partial decoupling'' property occurs for dynamical pumps that fall within the so-called ``Class~II'' 
designation of Ref.~\cite{Dienes:2023ziv}, with a transition time depending implicitly on the Hubble 
parameter $H(t)$ rather than $t$ itself.

\begin{figure}
\centering 
  \includegraphics[keepaspectratio, width=0.53\textwidth] {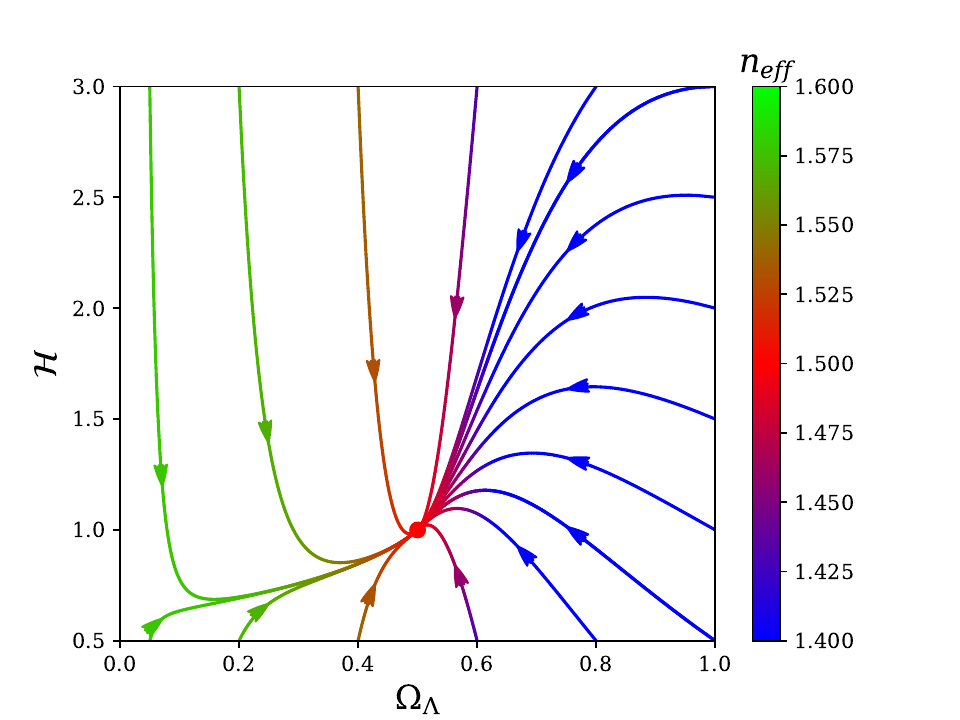}
\caption{
Attractor plot for a vacuum-energy/matter stasis with $\alpha\delta=1$ and $\gamma\delta=7$, 
demonstrating that our system dynamically flows to the stasis solution $(\Omega_\Lambda,\calH)=(1/2,1)$ 
for all initial values of $\Omega_\Lambda$ and $\calH$.  Just as in Fig.~\ref{fig:stasis}, these 
trajectories are plotted with colors indicating the corresponding values of $n_{\rm eff}$ that are 
realized at each moment during the time-evolution of our system.  Although $n_{\rm eff}\not = 3/2$ at 
all moments prior to reaching the stasis solution,  we observe that $n_{\rm eff}\to 3/2$ along every 
trajectory within the $(\Omega_\Lambda,\calH)$-plane as the fixed-point stasis solution is reached.
\label{fig:newattractorplot}}
\end{figure}

The differential equations in Eq.~(\ref{dyn}) allow us to map out the dynamical trajectories of this 
system.  Moreover, at any moment during this evolution we can evaluate $n_{\rm eff}$, obtaining
\beq
  n_{\rm eff} ~=~ \frac{3}{2} -\frac{1}{2} \frac{\partial_\Ncal \ln\Omega_\Lambda}{
    \partial_\Ncal \ln\rho_\Lambda} + \frac{\partial_\Ncal \ln(1+w_\Lambda \Omega_\Lambda)}
    {\partial_\Ncal \ln\rho_\Lambda}~.
\eeq
Indeed, this is the analogue of Eq.~(\ref{nefftrajectory}).  We thus obtain an attractor plot which is 
analogous to that in Fig.~\ref{fig:stasis}, only now plotted for our vacuum-energy/matter stasis.
This plot is shown in Fig.~\ref{fig:newattractorplot}.  As expected, this is essentially a version of 
the middle panel of Fig.~16 of Ref.~\cite{Dienes:2023ziv} except that the trajectories have been colored 
in order to indicate their corresponding values of $n_{\rm eff}$.   Once again, we see that 
$n_{\rm eff}\not=3/2$ at generic moments during the time-evolution of our system, but that 
$n_{\rm eff}\to 3/2$ along every possible trajectory as our system approaches the stasis solution.  
Unlike the situation in Fig.~\ref{fig:attractorTowerNeff}, however, we find that along each dynamical 
trajectory the value of $n_{\rm eff}$ either increases or decreases monotonically towards $3/2$.

A similar story exists for other tower-based stases.  For example, a stasis between vacuum energy and 
radiation was presented in Ref.~\cite{Dienes:2023ziv}, but the results for such a stasis are completely 
analogous to those presented here.  Likewise, it was also shown in Ref.~\cite{Dienes:2023ziv} that there 
can even be a {\it triple stasis}\/ during which vacuum energy, matter, and radiation can all simultaneously 
co-exist and remain in stasis with each other.  This result is somewhat non-trivial, involving the 
simultaneous operation of {\it two}\/ $n=1$ pumps, one of the form of Eq.~(\ref{pumpavgGamma2})
and the other of the form of Eq.~(\ref{phasetransitionpump}).  The important subtlety in such a case 
is that each of these pumps must now be capable of operating within a {\it three}\/-component cosmology 
which also includes a third energy component which is not directly involved in the pump action but which 
nevertheless affects the rate of cosmological expansion.   However, it was found in 
Ref.~\cite{Dienes:2023ziv} that this subtlety only requires the further restriction that $\gamma=1$ 
within the matter/radiation pump $P_{M\gamma}$.   

Within the analysis we have presented here, this restriction then implies that the scaling factor $u$ 
for the matter/radiation stasis in Eq.~(\ref{ucorrected}) must be equal to the scaling factor $u$ for 
the vacuum-energy stasis in Eq.~(\ref{ucorrected2}):
\beq
  u_{M\gamma} ~=~ u_{\Lambda M}~.
\eeq
Given the relation between these $u$-factors and the corresponding level-shift velocities $v$ in 
Eq.~(\ref{vu}), this in turn implies that
\beq
  v_{M\gamma} ~=~ v_{\Lambda M}~.
\eeq
More formally, this implies that our system has only a single global evolution operator $\calT$ in
Eq.~(\ref{shift_operations}), as required for a consistent triple stasis.

We close with two further comments.  First, we point out that there also exists a closely-related 
matter/radiation stasis in which the matter states within the tower are realized as primordial black
holes~\cite{Barrow:1991dn,Dienes:2022zgd}.  In such cases, the decay from matter to radiation occurs 
through Hawking radiation.  Ultimately, the underlying mathematics is similar to that described here, 
except that our decay transitions proceed {\it up}\/ (rather than down) the tower.  This change of 
sign for the shift velocity is ultimately immaterial for determining the effective $n$-value, and 
thus the arguments we have given here apply to this black-hole-based stasis as well.

Finally, as indicated in Eq.~(\ref{scalings2}), we remark that our analysis in this section has 
implicitly assumed that our tower has a mass spectrum in which the masses $m_\ell$ grow polynomially 
with $\ell$.   However, it has been shown~\cite{Halverson:2024oir} that stasis can emerge even in 
tower-based theories in which the masses $m_\ell$ grow {\it exponentially}\/ with $\ell$.  At first 
glance, it might seem that such a stasis would have fundamentally different characteristics. 
However, we have found that an analysis of such theories along the lines presented here ultimately 
reaches the same conclusions as for the theories involving polynomially growing towers, namely that 
$n_{\rm eff}\to 3/2$ as our system approaches stasis.


\section{A new look at thermal stasis:  How $n=2$ pumps produce $n_{\rm eff}=3/2$ behavior \label{sec:thermal}}


We now examine the $n=2$ thermal-stasis model of Ref.~{\cite{Barber:2024vui} and demonstrate that the 
value of $n_{\rm eff}$ in this model also experiences a deformation --- in this case a 
{\it thermally-induced}\/ deformation ---  to $n_{\rm eff}=3/2$.  As we shall see, our analysis in 
this case is actually far simpler than those in Sect.~\ref{sec:tower} because we no longer have the 
complications stemming from the existence of an entire tower of states.

\subsection{Recalling the thermal stasis model}

The model of thermal stasis that we shall consider~\cite{Barber:2024vui} assumes the existence of two 
cosmological species --- matter and radiation.  The matter, collectively denoted $M$, is represented 
as a single non-relativistic field of mass $m$ which collectively has energy density $\rho_M$, 
abundance $\Omega_M$, and equation-of-state parameter $w_M=0$.  Although we shall often refer to this 
energy component as being associated with matter, it is more specifically associated with the 
{\it rest-mass}\/ energy of our matter field.

We shall also assume that the matter particles that result from the excitations of this field are in 
thermal equilibrium with each other and together constitute an ideal gas of temperature $T$.  
This in turn implies that these matter particles also have non-zero kinetic energies.  Since kinetic 
energy is a distinct form of energy relative to rest-mass energy  --- and even has its own 
equation-of-state parameter $w_\KE= 2/3$ --- this means that we shall need to track the total kinetic 
energy of our $M$-particle gas as yet another energy component associated with our matter fields in the 
corresponding cosmology.  We shall therefore assume that the kinetic energy associated with our matter 
particles has an energy density $\rho_\KE$, abundance $\Omega_\KE$, and equation-of-state parameter 
$w_\KE= 2/3$.   Indeed, $\Omega_\KE$ is proportional to the temperature $T$.  Of course, our 
assumption that the matter particles are non-relativistic implies that $\Omega_\KE\ll \Omega_M$.

Finally, the second species in this model, collectively denoted $\chi$, is represented as a radiation 
field whose corresponding quantities are $\rho_\gamma$, $\Omega_\gamma$, and $w_\gamma= 1/3$.  The quanta 
associated with this field are assumed either to be completely massless or to have masses sufficiently 
small that these field quanta remain effectively relativistic throughout the stasis epoch.

Given these assumptions, we thus have a three-component universe whose different energy components are 
rest-mass energy with $w=0$, radiation with $w=1/3$, and kinetic energy with $w=2/3$.  The first and 
third of these are associated with the matter in our universe, while the second is associated with the 
radiation.  In such a universe, cosmological expansion tends to increase  $\Omega_M$ and decrease 
$\Omega_\KE$.  Likewise, given that $\Omega_\KE\ll \Omega_M$, cosmological expansion also tends to 
decrease $\Omega_\gamma$.  A stasis between $\Omega_M$ and $\Omega_\gamma$ --- one which we shall 
simply call a ``matter/radiation'' stasis --- can therefore emerge if there exists a counterbalancing 
process that converts rest-mass energy back to radiation.  

Even though we have only a single species of matter field (as opposed to an entire tower of such fields), 
it turns out that a simple $2\to 2$ {\it annihilation}\/ process of the form $MM\to \chi\chi$ can do the 
trick.  As familiar from studies of thermal freezeout, such a process corresponds to an $n=2$ pump which 
takes the form
\beq
  P_{M\gamma}^{(\rho)}~=~
    \frac{1}{m} \langle \sigma v\rangle \rho_M^2~
  \label{annihilation_pump}
\eeq
where $\langle \sigma v\rangle$ is the thermally averaged product of cross-section and relative velocity 
between two incoming matter particles.  Thanks to the thermal averaging of this product, this pump has an 
explicit dependence on the temperature of the gas of matter particles, making this an intrinsically thermal 
pump.  This in turn leads to an explicitly thermal stasis.

As in Ref.~\cite{Barber:2024vui}, we assume that this cross-section scales as
\beq
   \sigma v ~=~ C \,\left(
     \frac{ |\vec p_\CM|}{m}\right)^q
\eeq
where $C$ is an overall 
constant, where $\vec p_\CM$
is the 
three-momentum of either of the two incoming annihilating matter particles in the center-of-mass (CM) frame, 
and where $q$ is an arbitrary exponent.  For various technical reasons~\cite{Barber:2024vui}, we restrict 
our attention to values of $q$ within the range 
\beq
  -6 + 2\sqrt{3} ~\leq~ q ~\leq -3/2~.
  \label{q-range}
\eeq
It turns out that within this range, $q= -2$ is a particularly compelling value which can be directly 
realized in straightforward particle-physics models~\cite{Barber:2024vui}.

In the following we shall define the {\it coldness}
\beq
  \Xi~\equiv~ \frac{T^q \rho_M}{m^{q+4}}~.
\label{eq:DefOfSParam}
\eeq
Indeed, since $q<0$, we see that $\Xi$ increases when $T$ decreases.  Given this definition, we then
find~\cite{Barber:2024vui} that for any value of $q$ within the range in Eq.~(\ref{q-range}) this system 
develops a thermal matter/radiation stasis with
\beqn
  \barOmega_M ~&=&~ 1 + \frac{2q+3}{1+q^2/6} \nonumber\\
  \barXi ~&=&~ \frac{1}{\barOmega_M} 
    \left[ \frac{1-\barOmega_M}{\widehat C A(q)} \right]^2 ~
  \label{eq:stasisvalues}
\eeqn
where
\beq
  \widehat C ~\equiv~ \sqrt{\frac{3}{8\pi G}}\, m \,C~
  \label{Chatdef}
\eeq
and where 
\beq 
  A(q) ~\equiv~ \frac{2}{\sqrt{\pi}} \,
    \Gamma\left(\frac{q+3}{2}\right)~.
  \label{eq:Adef}
\eeq
It is further shown in Ref.~\cite{Barber:2024vui} that this stasis is a global attractor, and 
thus that this thermal system eventually reaches stasis regardless of its initial conditions.

It is noteworthy that this stasis does not involve a fixed temperature (or equivalently a fixed 
$\Omega_\KE$ abundance).  Instead, during this stasis it is the coldness $\Xi$ which approaches 
a fixed value.  By contrast, both $\rho_M$ and $T$ together drop to smaller and smaller values
in such a balanced way as to hold the coldness fixed at a non-trivial value $\barXi$ which depends 
on both $q$ and $\widehat C$.

\subsection{Obtaining $n_{\rm eff}=3/2$ from thermal effects\label{sec:n32thermal}}

In order to evaluate $n_{\rm eff}$ for this system, we begin
--- as in Sect.~\ref{sec:n32tower} ---
by evaluating $\partial_\calN P_{\Lambda M}^{(\rho)}$.
For this system, our pump is given by
\beqn
   P_{M\gamma}^{(\rho)} ~&=&~ \frac{1}{m} \,
   \langle \sigma v\rangle\, \rho_M^2
   \nonumber\\
   &=&~ \frac{C}{m} \,A(q)\, \left(\frac{T}{m}\right)^{q/2} \,\rho_M^2 \nonumber\\
   &=&~ C \,A(q)\, m \,
       \sqrt{\Xi} \,\rho_M^{3/2}~
\label{thermalpump}
\eeqn
where $\Xi$ is defined in Eq.~(\ref{eq:DefOfSParam}).   
This in turn implies that
\beq
\partial_\calN P_{M\gamma}^{(\rho)} ~=~
    \left(
\frac{3}{2} \frac{ \partial_\calN \rho_M}{\rho_M} + \frac{1}{2}
\frac{\partial_\calN \Xi}{\Xi}\right) 
   P_{M\gamma}^{(\rho)}
\eeq
or equivalently
\beq
\partial_\calN\log P_{M\gamma}^{(\rho)} ~=~
    \left(
\frac{3}{2} \, \partial_\calN \log \rho_M + \frac{1}{2}\,
\partial_\calN \log \Xi\right) 
  ~.~
\eeq
We therefore find that
\beq
n_{\rm eff} ~\equiv~
\frac{ \partial_\calN\log P_{M\gamma}^{(\rho)}}
{\partial_\calN \log \rho_M }
~=~
  \frac{3}{2} + \frac{1}{2}\,
  \frac{ \partial_\calN\log \Xi}
{\partial_\calN \log \rho_M }~.
\label{altform}
\eeq
However, during stasis, the coldness $\Xi$ is a constant.
Thus, we have
\beq 
  n_{\rm eff} ~=~ \frac{3}{2}~ 
\eeq
during stasis.  Indeed, we obtain this value of $n_{\rm eff}$ for all values of $q$.  From this perspective 
we see that the coldness variable $\Xi$ is special precisely because this is the quantity that remains fixed 
during stasis, allowing $n_{\rm eff}$ to take the form given in Eq.~(\ref{altform}).

\begin{figure}[t!]
    \centering
    \includegraphics[keepaspectratio, width=0.53\textwidth]{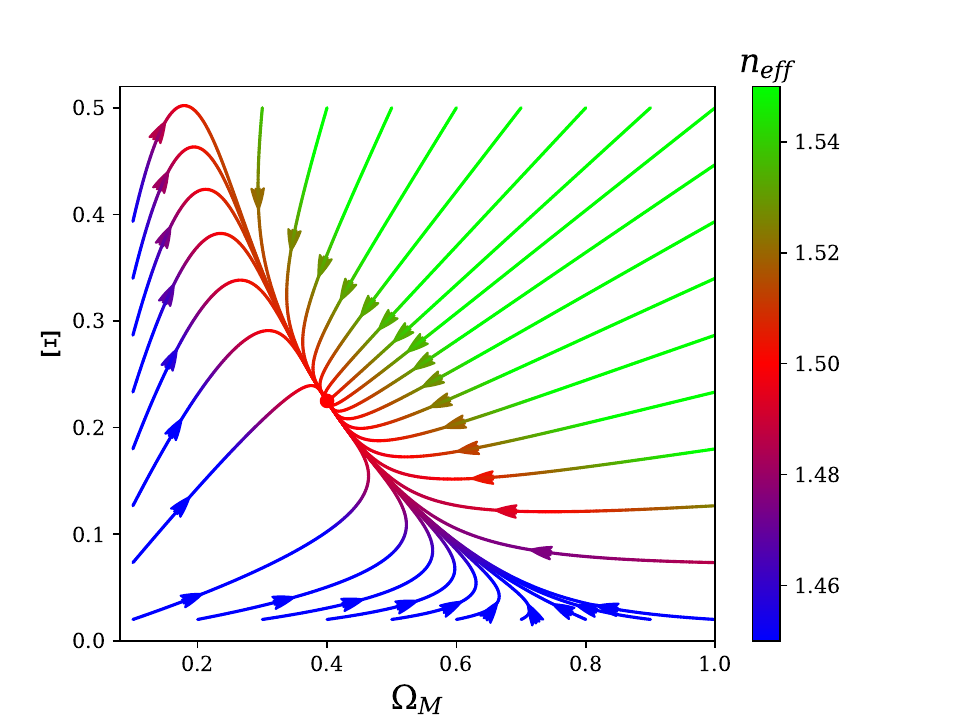}
    \caption{Attractor plot for thermal stasis, taking benchmark values $q=-2$ and $\widehat C =1$.
 These benchmark values correspond to $\barOmega_M=0.4$ and $\barXi= 9/40=0.225$ and lead to the same dynamical trajectories as in Fig.~4 of Ref.~\cite{Barber:2024vui}.~  
 Indeed, for all initial values of $\Omega_M$ and $\Xi$, these trajectories evolve towards the stasis solution.  However, in this figure these trajectories are plotted with colors indicating the corresponding values of $n_{\rm eff}$. 
 As expected, we find that $n_{\rm eff}\to 3/2$ in all cases as our system approaches stasis.
}
 \label{thermal_stasis_plot_2d}
\end{figure}

It is easy to recast this result in a way that explicitly demonstrates how this value of $n_{\rm eff}$ 
evolves due to thermal effects.  If we were to start instead from the second line of Eq.~(\ref{thermalpump}) 
and proceed as above, we would obtain
\beq
  n_{\rm eff} ~=~ 2 + \frac{q}{2} \,\frac{\partial_\calN \log T}{\partial_\calN  \log \rho_M}~.
  \label{thermaldeformation}
\eeq
With $n_{\rm eff}$ written in this form, we immediately see that it is the variation of the 
temperature $T$ --- even during stasis --- which injects an extra time-dependence into our pump beyond 
the time dependence implied by its dependence on $\rho_M$.  Indeed, the fact that $\partial_\calN \Xi=0$ 
during stasis immediately implies the stasis relation
\beq
  \partial_\calN \log T ~=~ -\frac{1}{q} \partial_\calN \log \rho_M ~.
\eeq
Inserting this into Eq.~(\ref{thermaldeformation}) then immediately leads to the result that
$n_{\rm eff} = 3/2$ during stasis.  Thus, we see that it is the continuously dropping temperature during 
the stasis epoch that is ultimately responsible for deforming $n_{\rm eff}$ to the stasis value 
$n_{\rm eff}=3/2$.  Indeed, this continually dropping temperature during thermal stasis 
plays the same role as the continually downward level-shifting plays during tower-based stases.

Just as with the other stases we have discussed, we can also examine how $n_{\rm eff}$ approaches $3/2$ 
as our dynamical system evolves towards stasis.  In general, the differential equations that govern the 
dynamical trajectories for this system are given by
\beqn
  \frac{d\Omega_M}{d\calN} ~&=&~ \Omega_M \left[1-\Omega_M 
    - \widehat{C} A(q) \sqrt{ \Xi\,\Omega_M} \, \right]\nonumber\\
  \frac{d\Xi}{d\calN} ~&=&~  \Xi\left[ -\left(2q+3\right)   
    -\widehat{C} \left(1+\frac{q^2}{6}\right) A(q) \sqrt{ \Xi \,\Omega_M} 
    \right]~.\nonumber\\   
  \label{Sweqs_expanded}
\eeqn
Moreover, for any point during the evolution of this system we can evaluate $n_{\rm eff}$ via 
Eq.~(\ref{altform}).  Taking the benchmark values $q= -2$ and $\hat C=1$ (implying $\barOmega= 0.4$ and 
$\barXi = 9/40=0.225$) then leads to the results shown in Fig.~\ref{thermal_stasis_plot_2d}.  Once again 
we observe that $n_{\rm eff}\to 3/2$ along all dynamical trajectories as our system approaches stasis.  
However, at more distant points, $n_{\rm eff}$ behaves non-monotonically.  In this connection, we remark 
that there exist minimum and maximum values of $n_{\rm eff}$ for any value of $q$:
\beq
  2 + \frac{q}{3} ~\leq~ n_{\rm eff} ~ \leq~ 2 + \frac{q^2}{12}~.
\eeq
Indeed, for $q= -2$ this yields
\beq
  \frac{4}{3}  ~\leq ~n_{\rm eff} ~\leq~ \frac{7}{3} ~.
\eeq       
These limiting values correspond to taking $\Xi \Omega_M\ll \overline{\Xi}\barOmega_M$ and 
$\Xi \Omega_M\gg \overline{\Xi}\barOmega_M$, respectively.


\section{Extracting the underlying $n=3/2$ theories \label{sec:extracting}}


Thus far, we have shown that each of our $n=1$ or $n=2$ theories evolves towards a stasis state 
which exhibits an $n_{\rm eff}=3/2$ behavior.  In this section, we shall push this observation 
one step further and demonstrate that each of these theories can actually be reformulated so as 
to have a manifestly $n=3/2$ structure right from the beginning.  In other words, by reshuffling and 
repackaging the  different degrees of freedom in these theories, we shall demonstrate that each of 
these theories can actually be reformulated in such a way that it has a pump of the same general form 
as that given for the $n=3/2$ theory given in Sect.~\ref{sec:threehalftheory}.~  This then demonstrates 
that each of these theories can itself be reformulated as an $n=3/2$ theory --- \ie, a theory which 
is similar to the theory we discussed in Sect.~\ref{sec:threehalftheory}.~  Note that this assertion 
concerns the theories themselves, and not merely their stasis solutions.  Thus this observation holds 
not only during stasis, but throughout the dynamical evolution of these theories.

In this context, it is important to note that what appears in Sect.~\ref{sec:threehalftheory} is not 
a single theory.   Certainly a unique theory is specified whenever a unique pump is specified.  However, 
in Sect.~\ref{sec:threehalftheory} we were content to analyze the properties that emerge when the 
pump $P_{ij}^{(\rho)}$ is taken to have the general form in Eq.~(\ref{genn}) with $n=3/2$.  In 
particular, we did not specify a particular value or expression for the coefficient $Z$.~  Of course, 
certain facts are known about $Z$ --- for example, it must be positive by convention,  so that 
$P_{ij}^{(\rho)}$ indeed involves the transfer of energy from the $i^{\rm th}$ energy component 
to the $j^{\rm th}$ component.  However, there is also another critical property that $Z$ must have. 
In Sect.~\ref{relationtopumptimescaling}, we demonstrated that any pump for which $n_{\rm eff}=3/2$ during 
stasis will exhibit the required $P\sim 1/t$ scaling during stasis.  Moreover, this will also be true 
for any pump of the form in Eq.~(\ref{genn}) with $n=3/2$ so long as $Z$ does not contribute any further 
time-scaling of its own during stasis.  We thus conclude that $Z$ must approach a constant during 
stasis for any pump of the form in Eq.~(\ref{genn}) with $n=3/2$.   This then becomes an additional 
requirement for $Z$.

However, even requiring these properties does not uniquely specify $Z$.  As a result, there is still a 
non-trivial set of possible $n=3/2$ theories, each with its own expression for $Z$ satisfying the above 
conditions.  In other words, the set of theories for which $Z$ exhibits these two properties constitutes 
an $n=3/2$ {\it equivalence class}\/ of theories.  What we are asserting, then, is that our analysis in
Sect.~\ref{sec:threehalftheory} actually applies to this entire $n=3/2$ equivalence class, 
and that each of the theories we have examined in Sects.~\ref{sec:tower} and \ref{sec:thermal} is a 
secretly a member of this equivalence class.  Of course, the theories in Sects.~\ref{sec:tower} and 
\ref{sec:thermal} are very special members of this $n=3/2$ equivalence class: they are theories which 
are originally formulated as $n=1$ or $n=2$ theories and thereby have explicit particle-physics 
representations in terms of underlying scattering/decay amplitudes or Feynman diagrams.

In order to demonstrate that these $n=1$ and $n=2$  theories are indeed members of this $n=3/2$ 
equivalence class, we must somehow algebraically rewrite these pumps so that they change from having 
$n=1$ or $n=2$ to having $n=3/2$.  In other words, we shall need to transform these pumps from what we 
shall call the ``Feynman picture'' to what we may call the ``stasis picture''.  However, simply rewriting 
these pumps so as to change their apparent values of $n$ is only part of the story.  

In order to understand the remaining elements, let us first take a bird's-eye view of the entire situation.
We began by specifying our theories by writing their pumps in the form given in Eq.~(\ref{genn}), where $n$ 
is the number of external in-state $\phi_i$ particles in the Feynman diagram that represents the 
energy-transfer process which underlies the pump.  However, in this picture we also implicitly demanded that 
$Z>0$ and also that $Z$ be $\rho_i$-independent, so that $n$ indeed represents the number of external 
in-state $\phi_i$ particles.  In other words, even though we allowed 
\beq 
  \partial_\calN Z ~\not= ~0~,
  \label{Zcond2}
\eeq
we nevertheless demanded
\beq 
  \partial_{\rho_i} Z ~=~ 0~.
  \label{Zcond1} 
\eeq
We shall refer to such a Feynman-diagram-motivated expression for the pump as constituting the 
``Feynman picture.''  Indeed, if this theory ultimately exhibits a stasis solution, then the extra 
time-dependence for $Z$ in Eq.~(\ref{Zcond2}) would be needed in order  to endow the entire pump 
with the required $P\sim 1/t$ scaling.

However, by shifting some number of factors of $\rho_i$ into or out of the coefficient $Z$ (\ie, by 
multiplying or dividing $Z$ by factors of $\rho_i$) along with other algebraic manipulations, 
we can algebraically recast this same pump within Eq.~(\ref{genn}) into a number of alternate forms
\beq 
  P^{(\rho)}_{ij} ~=~ Z' \rho_i^{n'} ~.
  \label{refo} 
\eeq 
Clearly many possible reformulations of this type are possible, each with its own unique value of $n'$.
However, if our underlying theory supports a stasis solution, then we shall be particularly interested 
in the particular reformulation for which $n'=3/2$.  Indeed, within this formulation, $Z'$ is positive.
Moreover, we know from Sect.~\ref{relationtopumptimescaling} that taking $n'=3/2$ is already sufficient 
to guarantee the proper overall pump scaling $P_{ij}\sim 1/t$.  It therefore follows that that $Z'$
must become constant during stasis.  Of course, since $n'\not= n$, we can no longer demand that $Z'$ be 
independent of $\rho_i$.  Thus, in this special primed formulation, we now allow
\beq 
  \partial_{\rho_i} Z' ~\not= ~ 0~,
  \label{Acond1} 
\eeq
but we instead demand
\beq 
     \partial_\calN Z' ~= ~0~.
\label{Acond2}
\eeq
Note that these conditions for $Z'$ are flipped relative to those for $Z$ in Eqs.~(\ref{Zcond2}) 
and~(\ref{Zcond1}).

We shall refer to this primed formulation with $n'=3/2$ and with $\partial_\calN Z'=0$ during stasis as 
the ``stasis picture'' --- it describes the same theory, but with the degrees of freedom reshuffled. 
Indeed, this difference concerning which quantities are time-dependent in the Feynman versus stasis 
pictures is completely analogous to the difference between the Schr\"odinger and Heisenberg formulations 
of quantum mechanics --- in the former it is the states that carry the time-dependence while in the 
latter this dependence is carried by the operators.  Indeed, in both cases the different pictures simply 
result from shifting the time-dependence from one set of variables to another.  Of course, such a stasis 
picture exists for only those theories that are ultimately capable of supporting stasis, and one of the 
results of this paper is that such a stasis picture necessarily has $n'=3/2$.  Otherwise, if our original 
particle-physics theory did not support a stasis, it would not be possible to recast its pump into a 
stasis picture.  Indeed, these conditions on $Z'$ would not be satisfied when such a theory is 
reformulated with $n' = 3/2$.

These observations also explain the behavior of $n_{\rm eff}$.  In general, we recall that $n_{\rm eff}$ 
is defined in Eq.~(\ref{ndef1}).  As such, $n_{\rm eff}$ depends on the time dependence of the 
{\it entire pump}\/.  This means that $n_{\rm eff}$ not only depends on $n$ or $n'$ (which indicate 
the number of $\rho_i$ factors in the pump), but also incorporates the time-dependence of the leading 
coefficients $Z$ or $Z'$.  We then find 
\beq
  n_{\rm eff} ~=~ n + \frac{\partial_\calN Z}{\partial_\calN \log \rho_i} ~=~
  n' + \frac{\partial_\calN Z'}{\partial_\calN \log \rho_i}~.
  \label{neffconstruction}
\eeq
However, the first of these equations is already familiar to us from previous sections, illustrating 
how our original values of $n$ in each case are ultimately ``deformed'' with additional contributions 
due to the time-dependence of $Z$ to produce $n_{\rm eff}=3/2$.  Moreover, we see from the second 
equation within Eq.~(\ref{neffconstruction}) that during stasis we have $\partial_\calN Z'=0$, which 
implies that $n_{\rm eff}= n'$ during stasis.  Of course, we have already seen that we must always 
have $n'=3/2$ within the stasis picture.  These observation then explain why $n_{\rm eff}$ flows to 
$3/2$ during stasis.

In the rest of this section, therefore, our goal will be to demonstrate that each of the $n=1$ or $n=2$ 
theories we have examined in Sects.~\ref{sec:tower} and \ref{sec:thermal} can indeed be recast into a
stasis picture.  This is equivalent to extracting  the hidden $n=3/2$ stasis theories corresponding to 
the pumps in question, and is tantamount to establishing that our $n=1$ or $n=2$ theories are indeed 
secretly members of this $n=3/2$ equivalence class.  For each of these $n=1$ or $n=2$ theories, we shall 
do this by explicitly performing  a series of algebraic manipulations in order to recast the corresponding 
particle-physics pump into the form in Eq.~(\ref{refo}) where $n'=3/2$ and where the $Z'$ coefficients 
are positive, with $\partial_\calN Z'=0$ during stasis.  Moreover, in each case we shall carry out this 
reformulation in a manner which does not assume stasis, but which holds completely generally.

\subsection{Tower-based matter/radiation stasis} 

We begin by considering the tower-based matter/radiation stasis which was discussed in 
Sects.~\ref{sect:towerbasedreview} through \ref{sec:n32tower}.~  As we recall, this stasis is apparently 
an $n=1$ stasis, with a pump given by $P^{(\rho)}_{M\gamma}= \langle \Gamma \rangle \rho_M$.  We 
therefore wish to rewrite this pump in the form given in Eq.~(\ref{refo}) with $n'=3/2$.

A sequence of algebraic manipulations that accomplishes this is given by
\beqn
  P^{(\rho)}_{M\gamma} &=& \langle \Gamma \rangle \rho_M\nonumber\\
    &= & \frac{\eta}{\gamma} \, \rho_M  \frac{1}{t} \nonumber\\
    &=&  \frac{\eta}{\gamma}  \rho_M \frac{3}{\barkappa} \frac{H}{\calH} \nonumber\\
    &=& \frac{3 \eta}{\gamma \barkappa}   \rho_M \sqrt{\frac{8\pi G}{3}} 
    \rho_{\rm tot}^{1/2} \frac{1}{\calH} \nonumber\\
    &=& \frac{3 \eta}{\gamma \barkappa} \sqrt{\frac{8\pi G}{3}}  \frac{1}{\Omega_M^{1/2} \calH} \rho_M^{3/2} ~.
  \label{manipulation1}
\eeqn
In performing this sequence of manipulations we have used Eq.~(\ref{bracketGamma}) in passing to the 
second line, we have recognized $\calH\equiv 3Ht/\barkappa$ in passing to the third line, and we have used 
the trivial relation $\rho_M=\Omega_M \rho_{\rm tot}$ in passing to the final line.  Note that each step of
Eq.~(\ref{manipulation1}) holds generally, independently of any stasis assumption.  Thus, for this theory, 
we have $Z=\langle \Gamma\rangle$ but
\beq
  Z' ~=~ \frac{3 \eta}{\gamma \barkappa} \sqrt{\frac{8\pi G}{3}}  \frac{1}{\Omega_M^{1/2} \calH}~.
\label{Z1}
\eeq
Indeed, this quantity is constant during stasis.

\subsection{Tower-based vacuum-energy/matter stasis}

In this case, our pump $P_{\Lambda M}^{(\rho)}(t)$ is the product of the density of states 
$n_{\hat t}(t)$ and a continuous variable $\widetilde \rho(t)$ which is {\it not}\/ the total 
vacuum-energy density $\rho_{\Lambda}$ but rather the differential energy density that is 
instantaneously being converted from vacuum energy to matter at the time $t$.  However, this 
theory also gives  rise to stasis, and thus --- even without assuming stasis --- this pump may 
also be reformulated into the stasis picture:
\beqn 
  P^{(\rho)}_{\Lambda M} &=& n_{\hat t} \widetilde\rho \nonumber\\ 
    &=& \left| \frac{d\ell}{d t_\ell}\right|_{t_\ell=t} \widetilde\rho \nonumber\\
    &= &  \frac{3\eta}{\kappa} H \rho_\Lambda  \nonumber\\
    &= & \frac{3\eta}{\kappa} \,
    \sqrt{\frac{8\pi G}{3}} \rho_{\rm tot}^{1/2} \,\rho_\Lambda \nonumber\\
    &= & 3\eta\, \sqrt{\frac{8\pi G}{3}} \frac{1}{\kappa\, \Omega_\Lambda^{1/2}} 
    \,\rho_\Lambda^{3/2}~ .
\label{tower_vac_pump_rewrite}
 \eeqn
Here the passage to the third line utilizes the alternate expression for this pump that was derived 
without assuming stasis in Eq.~(7.23) of Ref.~\cite{Dienes:2023ziv} --- an expression in which 
$\widetilde\rho$ is essentially replaced by $\rho_{\Lambda}$.  For this reason, it is the expression 
that appears on the third line which may be taken as the pump expressed in the ``Feynman picture''.   
By contrast, the passage to the final line utilizes the trivial relation 
$\rho_\Lambda= \Omega_\Lambda \rho_{\rm tot}$.

For this stasis we can look to the third line of Eq.~(\ref{tower_vac_pump_rewrite}) to identify 
$Z=3\eta H/\kappa$.  However, from the above analysis we see that
\begin{equation}
    Z' ~=~ 3\eta\,
\sqrt{\frac{8\pi G}{3}} \frac{1}{\kappa\, \Omega_\Lambda^{1/2}} ~.
\end{equation}
As a check, we observe that this quantity is indeed constant during stasis.

\subsection{Thermal stasis}

We now perform a similar transformation for the pump underlying thermal stasis: 
\beqn
  P^{(\rho)}_{M\gamma} &=&\frac{1}{m}\langle \sigma v\rangle \,\rho_M^2\nonumber\\
    &=&\frac{C}{m} A(q) \, T^{q/2}\, \rho_M^2\,\nonumber\\
    &=&  C m A(q)\,\Xi^{1/2}\,\rho_M^{3/2} ~.
  \label{thermal_rewrite} 
\eeqn
Thus, for this stasis, we have $Z= \langle \sigma v\rangle / m$ but 
\begin{equation}
  Z' = C m A(q)\, \Xi^{1/2}~.
  \label{thermal_rewrite2}
\end{equation}
This quantity is also constant during stasis.

In comparing the second and third lines of Eq.~(\ref{thermal_rewrite}), we learn that the coldness 
$\Xi$ is actually nothing but the ``deformed'' version of the temperature $T$, or equivalently the 
version of $T$ that is relevant in the stasis picture. Indeed, it is the coldness (and not $T$ itself) 
that  remains constant during thermal stasis.   This explains the importance of the coldness $\Xi$ in 
studies of thermal stasis.


\section{Discussion and conclusions}


In some sense, we have now come full circle.  We began by observing that taking $n=3/2$ in 
Eq.~(\ref{genn}) leads to the stasis described in Sect~\ref{sec:threehalftheory} in which the Hubble 
parameter is absent from the stasis condition in Eq.~(\ref{stasiscondition3}).   However, the 
non-integer value of $n$ appeared to preclude any particle-physics realization of such a stasis; 
indeed theories involving decay processes apparently lead to $n=1$ pumps while theories involving 
$2\to 2$ annihilation processes apparently lead to $n=2$ pumps.   It would thus seem that no 
particle-physics realizations of $n=3/2$ stasis are possible, since $n=3/2$ would appear to emerge 
from a process that would need to exist somewhere between decay and annihilation.

Given this, the rest of this paper can then be viewed as providing a gradual, incremental refutation 
of this conclusion.   First we proceeded to demonstrate that under time-evolution,  each of our $n=1$ 
and $n=2$ particle-physics theories is ultimately deformed in such a way that its effective $n$-value, 
as quantified through $n_{\rm eff}$, actually approaches the value $3/2$ as the theory approaches stasis, 
ultimately achieving $n_{\rm eff}=3/2$ during stasis.    We then demonstrated that in each case this is 
not simply a property of the stasis solution, but is instead a property of our particle-physics model 
itself, and that the pumps in these models can actually be recast into a form we dubbed the ``stasis 
picture'' in which the  $n=3/2$ behavior is manifest.  These models are therefore explicit examples of 
particle-physics models that --- despite initial appearances --- actually have $n=3/2$.   These examples 
are very different from each other --- some involve particle decay while others involve particle 
annihilation;  some are intrisically non-thermal while others are explicitly thermal;   and some 
involve entire towers of states while others have no towers at all.   Yet we have shown that they are 
all secretly $n=3/2$ theories. 
 
These results are perhaps even more remarkable when viewed in reverse.   If we had been assigned to 
construct a particle-physics realization of $n=3/2$ stasis, we would have started with a pump of the 
form $Z'\rho_i^{3/2}$.   We would then likely have attempted to construct an expression for $Z'$ which 
satisfies all of the $Z'$ requirements in Sect.~\ref{sec:extracting} and which can also be realized 
through standard particle-physics means.   {\it A priori}\/, this would have seemed a Herculean task.     
For example, in order to create a pump that transfers matter into radiation, one might start by examining 
what happens when a single matter species decays into radiation.  Indeed, such a quest would lead to a 
pump of the form $P_{M\gamma}\sim \Gamma \rho_M$ where $\Gamma$ is the width of the decaying particle 
and $\rho_M$ its energy density.  However, as we have noted earlier, a single decay width $\Gamma$ is 
typically a time-independent constant, which in turn implies that the pump $\Gamma \rho_M$ cannot 
exhibit an $n_{\rm eff}=3/2$ scaling.   Thus, one would conclude that such a theory based on a decaying 
particle cannot furnish a particle-physics representation of an $n=3/2$ stasis.  Indeed, it is non-trivial 
to recognize that one should perhaps consider {\it multiple}\/ decaying particles, whereupon the single 
decay width $\Gamma$ is replaced by the {\it average}\/ $\langle \Gamma \rangle$, and then to further 
imagine that an actual {\it tower}\/ of such matter states with particular scaling relations such as 
those in Eqs.~(\ref{scalings1}) and (\ref{scalings2}) could endow $\langle \Gamma\rangle$ with the 
missing $1/t$ scaling that would turn this into an $n=3/2$ system.   Yet this is precisely what occurs 
in the tower-based matter/radiation stasis we have examined in Sect.~\ref{sec:tower}.~   

In a similar vein, it is remarkable that our thermal stasis is also secretly an $n=3/2$ stasis.  As we 
have seen, this relies critically on the existence of a new thermodynamic quantity --- the so-called 
{\it coldness}\/ $\Xi$ --- which actually remains constant during stasis, even though the temperature 
itself is dropping throughout the stasis epoch.  Indeed, it is coldness, rather than temperature, that is 
the conserved quantity during stasis.  This is particularly evident within the passage from 
Eq.~(\ref{thermal_rewrite}) to Eq.~(\ref{thermal_rewrite2}): while the temperature $T$ is the critical 
variable in the Feynman picture, it is the coldness $\Xi$ that plays the analogous role in the stasis 
picture.  This suggests that coldness may play a more significant role in the thermodynamics of stasis 
than previously appreciated.

These observations also indicate possible pathways for constructing new models of stasis.  Recognizing 
that all stases must have an underlying $n=3/2$ structure, and given the above examples of 
particle-physics realizations of such $n=3/2$ stases, one might attempt to find new mechanisms for 
achieving the required time-dependence for $Z$.   For example, it is natural to imagine that within 
a particle-physics context the coefficient $Z$ would generally include not only the masses of the 
different particles involved in the relevant energy-transfer processes but also the couplings that 
govern their interactions.   Thus, one might imagine constructing theories of stasis in which these 
masses and couplings themselves each exhibit a non-trivial  time-dependence.   Our realizations 
concerning the required underlying $n=3/2$ structure can then serve as a guide as to what kinds of 
time-dependence might be required in order to yield a stasis solution.

As a final comment, we note that there might even be situations in which the Feynman picture --- as 
determined directly from an underlying Lagrangian --- {\it coincides}\/ with the stasis picture.  
In principle, this cannot arise in situations in which the pump corresponds to a single Feynman 
diagram, since such situations necessarily have $n\in \IZ$.  However, we can imagine other 
possibilities if multiple Feynman diagrams are involved.  For example,
let us consider a case in which two matter fields $A$ and $B$ interact with each other via 
the one-to-two process $A \leftrightarrow 2B$.  Let us further assume that the rates for 
both the forward and inverse processes are sufficiently high at early times (when the number 
densities of $A$ and $B$ are both large) that thermal equilibrium is maintained between 
$A$ and $B$.  If the energy densities of these non-relativistic species satisfy 
$\rho_A \gg \rho_B$ while in this equilibrium state, it then follows that 
$\rho_B\sim \rho_A^{1/2}$.  Thus, if there also exists a slower scattering process 
of the form $A B \rightarrow \gamma \gamma$, where $\gamma$ denotes a relativistic 
particle species functioning as radiation, then the pump associated with this process 
would scale as $P^{(\rho)} \sim \rho_A\rho_B\sim \rho_A^{3/2}$ while $A$ and $B$ 
remain in equilibrium.  This then exhibits the desired $n=3/2$ behavior.

Unfortunately, this model fails to give rise to stasis for a simple (and ultimately model-independent) 
reason:  as a consequence of conservation of energy, the equilibrium between $A$ and $B$ is inevitably 
temperature-dependent.  In general, such temperature dependence introduces  an additional time-dependence 
into the pump beyond that arising from its $n=3/2$ structure, thereby destabilizing the stasis that 
would otherwise emerge.  These sorts of difficulties make it difficult to construct a model that gives 
rise to the kind of $n=3/2$ pump that can produce a stable stasis epoch directly in the Feynman picture.
However, other more complex scenarios of this type may be possible.

We see, then, that all successful models of stasis share a common underlying feature:   they are all 
different manifestations of an $n=3/2$ stasis.  This provides us with a deeper understanding of the 
stasis phenomenon, and how it emerges from a wide variety of well-known and well-motivated models of 
BSM particle physics.  This also allows us to relate the different existing models of stasis to each other, 
and thereby compare the common roles that their different physical quantities play within the stasis 
environment.  But finally, and perhaps most importantly, these insights can potentially help point the 
way to new models of stasis which have not yet been realized.


\begin{acknowledgments}


We are happy to thank L.~Heurtier, F.~Huang, and T.\,M.\,P.~Tait for discussions.
The research activities of JB and KRD are supported in part by the U.S.\ Department 
of Energy under Grant DE-FG02-13ER41976 / DE-SC0009913; the research activities of 
KRD are also supported in part by the U.S.\ National Science Foundation through its 
employee IR/D program.  The research activities of BT are supported in part by the 
U.S.\ National Science Foundation under Grant PHY-2310622.  The opinions and conclusions
expressed herein are those of the authors, and do not represent any funding agencies. 

\end{acknowledgments}

\bibliography{TheLiterature2}

\begin{thebibliography}{9}%
\makeatletter
\providecommand \@ifxundefined [1]{%
 \@ifx{#1\undefined}
}%
\providecommand \@ifnum [1]{%
 \ifnum #1\expandafter \@firstoftwo
 \else \expandafter \@secondoftwo
 \fi
}%
\providecommand \@ifx [1]{%
 \ifx #1\expandafter \@firstoftwo
 \else \expandafter \@secondoftwo
 \fi
}%
\providecommand \natexlab [1]{#1}%
\providecommand \enquote  [1]{``#1''}%
\providecommand \bibnamefont  [1]{#1}%
\providecommand \bibfnamefont [1]{#1}%
\providecommand \citenamefont [1]{#1}%
\providecommand \href@noop [0]{\@secondoftwo}%
\providecommand \href [0]{\begingroup \@sanitize@url \@href}%
\providecommand \@href[1]{\@@startlink{#1}\@@href}%
\providecommand \@@href[1]{\endgroup#1\@@endlink}%
\providecommand \@sanitize@url [0]{\catcode `\\12\catcode `\$12\catcode
  `\&12\catcode `\#12\catcode `\^12\catcode `\_12\catcode `\%12\relax}%
\providecommand \@@startlink[1]{}%
\providecommand \@@endlink[0]{}%
\providecommand \url  [0]{\begingroup\@sanitize@url \@url }%
\providecommand \@url [1]{\endgroup\@href {#1}{\urlprefix }}%
\providecommand \urlprefix  [0]{URL }%
\providecommand \Eprint [0]{\href }%
\providecommand \doibase [0]{http://dx.doi.org/}%
\providecommand \selectlanguage [0]{\@gobble}%
\providecommand \bibinfo  [0]{\@secondoftwo}%
\providecommand \bibfield  [0]{\@secondoftwo}%
\providecommand \translation [1]{[#1]}%
\providecommand \BibitemOpen [0]{}%
\providecommand \bibitemStop [0]{}%
\providecommand \bibitemNoStop [0]{.\EOS\space}%
\providecommand \EOS [0]{\spacefactor3000\relax}%
\providecommand \BibitemShut  [1]{\csname bibitem#1\endcsname}%
\let\auto@bib@innerbib\@empty
\bibitem [{\citenamefont {Dienes}\ \emph
  {et~al.}(2022{\natexlab{a}})\citenamefont {Dienes}, \citenamefont {Heurtier},
  \citenamefont {Huang}, \citenamefont {Kim}, \citenamefont {Tait},\ and\
  \citenamefont {Thomas}}]{Dienes:2021woi}%
  \BibitemOpen
  \bibfield  {author} {\bibinfo {author} {\bibfnamefont {K.~R.}\ \bibnamefont
  {Dienes}}, \bibinfo {author} {\bibfnamefont {L.}~\bibnamefont {Heurtier}},
  \bibinfo {author} {\bibfnamefont {F.}~\bibnamefont {Huang}}, \bibinfo
  {author} {\bibfnamefont {D.}~\bibnamefont {Kim}}, \bibinfo {author}
  {\bibfnamefont {T.~M.~P.}\ \bibnamefont {Tait}}, \ and\ \bibinfo {author}
  {\bibfnamefont {B.}~\bibnamefont {Thomas}},\ }\href {\doibase
  10.1103/PhysRevD.105.023530} {\bibfield  {journal} {\bibinfo  {journal}
  {Phys. Rev. D}\ }\textbf {\bibinfo {volume} {105}},\ \bibinfo {pages}
  {023530} (\bibinfo {year} {2022}{\natexlab{a}})},\ \Eprint
  {http://arxiv.org/abs/2111.04753} {arXiv:2111.04753 [astro-ph.CO]}
  \BibitemShut {NoStop}%
\bibitem [{\citenamefont {Dienes}\ \emph
  {et~al.}(2024{\natexlab{a}})\citenamefont {Dienes}, \citenamefont {Heurtier},
  \citenamefont {Huang}, \citenamefont {Tait},\ and\ \citenamefont
  {Thomas}}]{Dienes:2023ziv}%
  \BibitemOpen
  \bibfield  {author} {\bibinfo {author} {\bibfnamefont {K.~R.}\ \bibnamefont
  {Dienes}}, \bibinfo {author} {\bibfnamefont {L.}~\bibnamefont {Heurtier}},
  \bibinfo {author} {\bibfnamefont {F.}~\bibnamefont {Huang}}, \bibinfo
  {author} {\bibfnamefont {T.~M.~P.}\ \bibnamefont {Tait}}, \ and\ \bibinfo
  {author} {\bibfnamefont {B.}~\bibnamefont {Thomas}},\ }\href {\doibase
  10.1103/PhysRevD.109.083508} {\bibfield  {journal} {\bibinfo  {journal}
  {Phys. Rev. D}\ }\textbf {\bibinfo {volume} {109}},\ \bibinfo {pages}
  {083508} (\bibinfo {year} {2024}{\natexlab{a}})},\ \Eprint
  {http://arxiv.org/abs/2309.10345} {arXiv:2309.10345 [astro-ph.CO]}
  \BibitemShut {NoStop}%
\bibitem [{\citenamefont {Dienes}\ \emph
  {et~al.}(2024{\natexlab{b}})\citenamefont {Dienes}, \citenamefont {Heurtier},
  \citenamefont {Huang}, \citenamefont {Tait},\ and\ \citenamefont
  {Thomas}}]{Dienes:2024wnu}%
  \BibitemOpen
  \bibfield  {author} {\bibinfo {author} {\bibfnamefont {K.~R.}\ \bibnamefont
  {Dienes}}, \bibinfo {author} {\bibfnamefont {L.}~\bibnamefont {Heurtier}},
  \bibinfo {author} {\bibfnamefont {F.}~\bibnamefont {Huang}}, \bibinfo
  {author} {\bibfnamefont {T.~M.~P.}\ \bibnamefont {Tait}}, \ and\ \bibinfo
  {author} {\bibfnamefont {B.}~\bibnamefont {Thomas}},\ }\href {\doibase
  10.1103/PhysRevD.110.123514} {\bibfield  {journal} {\bibinfo  {journal}
  {Phys. Rev. D}\ }\textbf {\bibinfo {volume} {110}},\ \bibinfo {pages}
  {123514} (\bibinfo {year} {2024}{\natexlab{b}})},\ \Eprint
  {http://arxiv.org/abs/2406.06830} {arXiv:2406.06830 [astro-ph.CO]}
  \BibitemShut {NoStop}%
\bibitem [{\citenamefont {Barrow}\ \emph {et~al.}(1991)\citenamefont {Barrow},
  \citenamefont {Copeland},\ and\ \citenamefont {Liddle}}]{Barrow:1991dn}%
  \BibitemOpen
  \bibfield  {author} {\bibinfo {author} {\bibfnamefont {J.~D.}\ \bibnamefont
  {Barrow}}, \bibinfo {author} {\bibfnamefont {E.~J.}\ \bibnamefont
  {Copeland}}, \ and\ \bibinfo {author} {\bibfnamefont {A.~R.}\ \bibnamefont
  {Liddle}},\ }\href@noop {} {\bibfield  {journal} {\bibinfo  {journal} {Mon.
  Not. Roy. Astron. Soc.}\ }\textbf {\bibinfo {volume} {253}},\ \bibinfo
  {pages} {675} (\bibinfo {year} {1991})}\BibitemShut {NoStop}%
\bibitem [{\citenamefont {Dienes}\ \emph
  {et~al.}(2022{\natexlab{b}})\citenamefont {Dienes}, \citenamefont {Heurtier},
  \citenamefont {Huang}, \citenamefont {Kim}, \citenamefont {Tait},\ and\
  \citenamefont {Thomas}}]{Dienes:2022zgd}%
  \BibitemOpen
  \bibfield  {author} {\bibinfo {author} {\bibfnamefont {K.~R.}\ \bibnamefont
  {Dienes}}, \bibinfo {author} {\bibfnamefont {L.}~\bibnamefont {Heurtier}},
  \bibinfo {author} {\bibfnamefont {F.}~\bibnamefont {Huang}}, \bibinfo
  {author} {\bibfnamefont {D.}~\bibnamefont {Kim}}, \bibinfo {author}
  {\bibfnamefont {T.~M.~P.}\ \bibnamefont {Tait}}, \ and\ \bibinfo {author}
  {\bibfnamefont {B.}~\bibnamefont {Thomas}},\ }\href@noop {} {\  (\bibinfo
  {year} {2022}{\natexlab{b}})},\ \Eprint {http://arxiv.org/abs/2212.01369}
  {arXiv:2212.01369 [astro-ph.CO]} \BibitemShut {NoStop}%
\bibitem [{\citenamefont {Halverson}\ and\ \citenamefont
  {Pandya}(2024)}]{Halverson:2024oir}%
  \BibitemOpen
  \bibfield  {author} {\bibinfo {author} {\bibfnamefont {J.}~\bibnamefont
  {Halverson}}\ and\ \bibinfo {author} {\bibfnamefont {S.}~\bibnamefont
  {Pandya}},\ }\href {\doibase 10.1103/PhysRevD.110.075041} {\bibfield
  {journal} {\bibinfo  {journal} {Phys. Rev. D}\ }\textbf {\bibinfo {volume}
  {110}},\ \bibinfo {pages} {075041} (\bibinfo {year} {2024})},\ \Eprint
  {http://arxiv.org/abs/2408.00835} {arXiv:2408.00835 [astro-ph.CO]}
  \BibitemShut {NoStop}%
\bibitem [{\citenamefont {Barber}\ \emph
  {et~al.}(2024{\natexlab{a}})\citenamefont {Barber}, \citenamefont {Dienes},\
  and\ \citenamefont {Thomas}}]{Barber:2024vui}%
  \BibitemOpen
  \bibfield  {author} {\bibinfo {author} {\bibfnamefont {J.}~\bibnamefont
  {Barber}}, \bibinfo {author} {\bibfnamefont {K.~R.}\ \bibnamefont {Dienes}},
  \ and\ \bibinfo {author} {\bibfnamefont {B.}~\bibnamefont {Thomas}},\ }\href
  {\doibase 10.1103/PhysRevD.110.123515} {\bibfield  {journal} {\bibinfo
  {journal} {Phys. Rev. D}\ }\textbf {\bibinfo {volume} {110}},\ \bibinfo
  {pages} {123515} (\bibinfo {year} {2024}{\natexlab{a}})},\ \Eprint
  {http://arxiv.org/abs/2408.16255} {arXiv:2408.16255 [astro-ph.CO]}
  \BibitemShut {NoStop}%
\bibitem [{\citenamefont {Barber}\ \emph
  {et~al.}(2024{\natexlab{b}})\citenamefont {Barber}, \citenamefont {Dienes},\
  and\ \citenamefont {Thomas}}]{Barber_toappear_model}%
  \BibitemOpen
  \bibfield  {author} {\bibinfo {author} {\bibfnamefont {J.}~\bibnamefont
  {Barber}}, \bibinfo {author} {\bibfnamefont {K.~R.}\ \bibnamefont {Dienes}},
  \ and\ \bibinfo {author} {\bibfnamefont {B.}~\bibnamefont {Thomas}},\
  }\href@noop {} {\  (\bibinfo {year} {2024}{\natexlab{b}})},\ \Eprint
  {http://arxiv.org/abs/Model of a Thermal Stasis Attractor, to appear} {Model
  of a Thermal Stasis Attractor, to appear} \BibitemShut {NoStop}%
\bibitem [{\citenamefont {Batell}\ \emph {et~al.}(2024)\citenamefont {Batell}
  \emph {et~al.}}]{Batell:2024dsi}%
  \BibitemOpen
  \bibfield  {author} {\bibinfo {author} {\bibfnamefont {B.}~\bibnamefont
  {Batell}} \emph {et~al.},\ }\href@noop {} {\  (\bibinfo {year} {2024})},\
  \Eprint {http://arxiv.org/abs/2411.04780} {arXiv:2411.04780 [astro-ph.CO]}
  \BibitemShut {NoStop}%
\end{thebibliography}%

\end{document}